\documentclass[11pt]{article}
\usepackage[dvips]{graphicx}

[1999/03/29 PSNFSS v.7.2
Times font as default roman
: S Rahtz]

\newcommand{\be}{\begin{equation}}
\newcommand{\ee}{\end{equation}}
\newcommand{\bea}{\begin{eqnarray}}
\newcommand{\eea}{\end{eqnarray}}
\newenvironment{technical}{\begin{quotation}\small}{\end{quotation}}

\begin{document}

\title{Imitation and contrarian behavior:\\ hyperbolic bubbles, 
crashes and chaos}
\thispagestyle{empty}

\author{A. Corcos${}^1$, J.-P. Eckmann${}^{2,3}$, A. Malaspinas${}^2$,
Y. Malevergne$^{4,5}$ and D. Sornette$^{4,6}$\\
$^1$CRIISEA,
Universit\'e de Picardie, BP 2716
80027 Amiens, France\\
$^2$D\'ept.~de Physique Th\'eorique, Universit\'e de Gen\`eve,
CH-1211 Gen\`eve 4, Switzerland\\
$^3$Section de Math\'ematiques, Universit\'e de Gen\`eve,
CH-1211 Gen\`eve 4, Switzerland\\
$^4$ Laboratoire de Physique de la Mati\`{e}re Condens\'{e}e\\ CNRS UMR6622 and
Universit\'{e} de Nice-Sophia Antipolis\\ B.P. 71, Parc
Valrose, 06108 Nice Cedex 2, France\\
$^5$ ISFA, University of Lyon 1, France\\
$^6$ Institute of Geophysics and
Planetary Physics and Department of Earth and Space Science\\
University of California, Los Angeles, California 90095
}

\date{\today}
\maketitle

\abstract{Imitative and contrarian behaviors are the two typical opposite attitudes
of investors in stock markets. 
We introduce a simple model to investigate their interplay
in a stock market where agents can take only two states, bullish or bearish.
Each bullish (bearish) agent polls $m$ ``friends'' and changes her opinion to bearish
(bullish)
if (1) at least $m \rho_{hb}$ ($m \rho_{bh}$)  among the $m$ agents inspected are bearish
(bullish) or  (2) at least  $m \rho_{hh}>m \rho_{hb}$ ($m \rho_{bb}>m \rho_{bh}$) 
among the $m$ agents inspected are bullish (bearish). The condition (1) (resp.~(2))
corresponds to imitative (resp.~antagonistic) behavior. In the limit where the number 
$N$ of agents is infinite, the dynamics of the fraction of bullish agents 
is deterministic and exhibits chaotic behavior in a
significant domain of the parameter space $\{\rho_{hb}, \rho_{bh}, \rho_{hh}, \rho_{bb}, m\}$.
A typical chaotic trajectory is characterized by intermittent phases of chaos, quasi-periodic
behavior and super-exponentially growing bubbles followed by crashes. A typical bubble
starts initially by growing at an exponential rate and then crosses over to a nonlinear
power law growth rate leading to a finite-time singularity. 
The reinjection mechanism provided by the contrarian behavior introduces a finite-size
effect, rounding off these singularities and leads to chaos. 
We document the main stylized facts
of this model in the symmetric and asymmetric cases. This model is one of the rare
agent-based models that give rise to interesting non-periodic complex
dynamics in the ``thermodynamic'' limit (of an infinite number $N$ of agents).
We also discuss the case of a finite number of agents, which introduces
an endogenous source of noise superimposed on the chaotic dynamics.
}

\thispagestyle{empty}
\pagenumbering{arabic}
\newpage
\setcounter{page}{1}

\begin{technical}
``{\it Human behavior is a main factor in how markets act. Indeed, sometimes markets 
act quickly, violently with little warning. [$\dots$]
Ultimately, history tells us that there will be a correction of some significant
dimension. I have no doubt that, human nature being what it is, that it is going
to happen again and again.}''  Alan Greenspan, Chairman of the Federal Reserve of the USA,
before the Committee on Banking and Financial Services, U.S. House of
Representatives, July 24, 1998.
\end{technical}

\section{Introduction}

In recent economic and finance research, there is a growing interest in 
incorporating ideas from social sciences to account for the fact that  
markets reflect the thoughts, emotions, and actions of real 
people as opposed to the idealized economic investor whose behavior
underlies the efficient market and random walk hypothesis. 
This was captured by the now famous pronouncement of Keynes 
(1936) that most investors' decisions ``{\em can only be taken
as a result of animal spirits -- of a spontaneous urge to action rather
than inaction, and not the outcome of a weighed average of benefits 
multiplied by the quantitative probabilities}''. A real
investor may intend to be rational and may try to optimize his actions, but
that rationality tends to be
hampered by cognitive biases, emotional quirks, and social influences.
``Behavioral finance'' 
is a growing research field (Thaler (1993), De Bondt and Thaler (1995), 
Shefrin (2000), Shleifer (2000),  Goldberg and von Nitzsch (2001)), which uses
psychology, sociology, and other behavioral theories to attempt
to explain the behavior of investors and money managers. The behavior
of financial markets is thought to result from varying
attitudes toward risk, the heterogeneity in the framing of information,
from cognitive errors, self-control and lack thereof,
from regret in financial decision-making, and from the influence of 
mass psychology.
Assumptions about the frailty of human rationality and the acceptance of such
drives as fear and greed are underlying the recipes developed over decades
by so-called technical analysts.

There is growing empirical evidence for the existence of herd or ``crowd'' 
behavior in speculative markets 
(Arthur (1987),  Bikhchandani {\it et al.} (1992), Johansen {\it et al.}(1999, 2000),
Orl\'ean (1986, 1990, 1992), Shiller (1984, 2000), Topol (1991), West (1988)). 
Herd behavior is often said to occur when many people take the same action,
because some mimic the actions of others. Herding
has been linked to many economic activities, such as investment recommendations
(Graham and Dodd  (1934), Scharfstein and Stein (1990)), 
price behavior of IPO's  (Initial Public Offering) (Welch (1992)) 
fads and customs (Bikhchandani {\it et al.} (1992)), earnings forecasts
(Trueman (1994)), corporate conservatism (Zwiebel (1995)) and delegated
portfolio management (Maug and Naik (1995)).

Here, we introduce arguably the simplest model capturing the interplay between
mimetic and contrarian behavior in a population of $N$ agents taking only two
possible states, ``bullish'' or ``bearish'' (buying or selling).
In the limit of an infinite number $N \to \infty$ of agents, the key 
variable which is the
fraction $p$ of bullish agents follows a chaotic deterministic dynamics
on a subspace of positive measure in the parameter space. 
Before explaining and analyzing the model in subsequent sections, we
compare it in three respects to standard theories of economic behavior.

{\bf 1.} Since in the limit $N \to \infty$,
the model operates on a purely deterministic basis, it
actually challenges the purely
external and unpredictable origin of market prices. Our model exploits the
continuous mimicry of financial markets to show that the disordered and random
aspect of the time series of prices can be in part explained not only 
by the advent of
``random'' news and events, but can also
be generated by the behavior of the agents fixing the prices.

In the limit $N \to \infty$, the dynamics of prices in our
model is deterministic and derives from the
theory of chaotic dynamical systems, which have the feature of exhibiting
endogenously perturbed motion. After the first papers on the theory of
chaotic systems, such as Lorenz (1963),
May (1976), (see, {\it e.g.}, Collet-Eckmann (1980)
for an early exposition),
a series of economic papers dealt with models mostly of 
growth---Benhabib-Day (1981),
Day (1982, 1983), Stutzer (1980).
Later, a vast and varied number of fields of economics were analyzed in the light
of the theory of chaos---Grandmont (1985, 1987),
Grandmont-Malgrange (1986). They extend from
macro-economics---business
cycles, models of class struggles, political economy---to 
micro-economics---models with
overlapping generations, optimizing behavior---and touch subjects
such as game theory and the theory of finance.
The applicability of these theories has been thoroughly tested on the stock
market prices---Brock {\it et al.}(1987),
Brock (1988),
Brock-Dechert (1988),
LeBaron (1988),
Brock {\it et al.}(1991),
Hsieh (1989),
Scheinkman-LeBaron (1989a, 1989b)---in studies which tried to detect signs
of non-linear effects and to nail down the deterministic nature of 
these prices. While the
theoretical models---Van Der Ploeg (1986),
De Grauwe-Vansanten (1990),
De Grauwe {\it et al.}(1993)---seem to agree
on the relevance of chaotic deterministic dynamics, the empirical
studies---Eckmann {\it et al.}(1988),
Hsieh-LeBaron (1988),
Hsieh (1989, 1991, 1992),
LeBaron (1988),
Scheinkman-LeBaron (1989a,b)---are less clear-cut, mostly because of lack of
sufficiently long time series (Eckmann-Ruelle (1992)),
or, because the
deterministic component of market behavior is necessarily overshadowed by the
inevitable external effects. An additional source of ``noise'' is found to
result from the finiteness of the number $N$ of agents. For
finite $N$, the deterministically
chaotic dynamics of the price is replaced by a stochastic dynamics shadowing 
the corresponding trajectories obtained for $N \to \infty$.

The model presented here shows a mechanism of price
fixing---decisions to buy or sell dictated by comparison with other
agents---which is at the origin of an instability of prices.
 From one period to the next, and in the absence of information other than the
anticipations of other agents, prices can continuously exhibit erratic behavior
and never stabilize, without diverging.
Thus, the model questions the fundamental hypothesis that equilibrium prices
have to converge to the intrinsic value of an asset.

{\bf 2.} We can also consider our model in the context of the
increasing market volatility of
financial markets. The volatility of prices generated by our chaotic model
could give a beginning of an explanation of the excess volatility observed on
financial markets---Grossman-Shiller (1981),
Fama (1965),
Flavin (1983),
Shiller (1981),
West (1988)---which traditional models, such as ARCH, try to
incorporate (Engle (1982),
Bollerslev {\it et al.}(1991),
Bollerslev (1987)).

{\bf 3.} Finally, we can see speculative bubbles in our model as a
natural consequence
of mimetism. We can compare this to the two basic trends in explaining the
problem of bubbles. The first makes reference to rational
anticipations---Muth (1961)---and rests on the hypothesis of efficient markets.
With fixed information, and knowing the dynamics of prices, the recurrence
relation for the price is seen to depend on the fundamental value and a
self-referential component, which tends to cause a deviation from the
fundamental value: this is a speculative bubble---Blanchard-Watson (1982).
This theory of rational speculative bubbles fails to explain the
birth of such events, and even less their collapse, which it does not predict
either. Recent developments
improve on these traditional approached by combining the
rational agents in the economy with irrational ``noise'' traders
(Johansen {\it et al.} (1999, 2000), Sornette and A. Johansen (2001)).
These noise traders are
imitative investors who reside on an interaction network. Neighbors of
an agent on this network can be viewed as the agent's friends or contacts, and
an agent will incorporate his neighbors' views regarding the stock into
his own view. These noise traders are responsible for triggering crashes. Sornette
and Andersen (2001) develop a similar model in which the noise traders induce 
a nonlinear positive feedback in the stock price dynamics with
an interplay between nonlinearity and multiplicative noise. The derived
hyperbolic stochastic finite-time singularity formula transforms a Gaussian
white noise into a rich time series possessing
all the stylized facts of empirical prices, as well as
accelerated speculative bubbles preceding crashes.

The second trend purports to explain speculative bubbles by a limitation of
rationality---Shiller (1984, 2000),
West (1988),
Topol (1991).
It allows to incorporate notions which the neo-classical analysis does not take
into account: asymmetry of information, inefficiency of prices, heterogeneity
of anticipations---Grossman (1977),
Grossman-Stiglitz (1980),
Grossman (1981),
Radner (1972, 1979).
In our approach, which follows the second trend, the agents act without
knowing the actual effect of their behavior: this contrasts the position of a
model-builder---Orl\'ean (1986, 1989, 1990, 1992). This, in turn, can 
lead to prices
which disconnect from the fundamental indicators of economics.

In the present paper we show that self-referred
behavior in financial markets can generate chaos and speculative bubbles.
They will be seen to be caused by mimetic behavior: bubbles will form due
to imitative behavior and
collapse when certain agents believe in the advent of a turn of trend, while
they observe the behavior of their peers.

Section 2 defines the model. Section 3 provides a qualitative understanding
and analysis of its dynamical properties. Section 4 extends it
with a quantitative analysis of the phases of speculative bubbles
in the symmetric case. Section 5 describes the statistical properties
of the price returns derived from its dynamics in the symmetric case.
Section 6 discusses the asymmetric case. Section 7 explores some effects
introduced by the finiteness $N<\infty$ of the number of agents.
In Section 8 we summarize our conclusions.

\section{The model}

We consider an economy in which the population makes choices between
two possible states when tomorrow's price is uncertain. The choice
depends on expected capital gains. The portfolio choice then becomes a
price expectation problem.
Each agent has a different set of informations, obtained by
observing other agents. Agents do not operate with reference to fundamental
value, but rather with respect to
{\it expected market price}. They are able to make profits
if their expectations are judiciously chosen. It is rational for the
agent---Keynes (1936),
Orl\'ean (1986, 1989), Sornette (2001) (see Chap. 4)---to take into account collective
judgments in order to make portfolio profits.

That is why, for constituting expectations at time $t+1$, the information
used by an
agent is the price expectation, at time $t$, of a certain sample of {\it
other} agents randomly chosen among the population.
This takes into account collective opinion and its expected correctness,
that is, their confidence (or absence thereof) in the continuation of 
a deviation
from the fundamental value. Their opinion refers to two kinds of price,
market price and fundamental value, as exhibited by Keynes (1936):

\item{1.} Speculation relying on short term action and especially market
opinion
and market price. The most important aspect is the market price expectation,
that is, the collective opinion about future market prices.
\item{2.} Firm behavior: long term behavior relying on economic reality
and fundamental value. This leads agents to detect excessive increase or
decrease of
market price and thus leads to anticipatory adaptation of the market 
price. This causes the
collapse of the bubble.

The importance of the interplay of these two classes of investing
(which can be used by a same agent alternatively), corresponding to
fundamental value investors and technical analysts (or trend followers),
has been stressed by several recent works (Lux and Marchesi (1999),
Farmer and Joshi, 2001) to
be essential in order to retrieve the important stylized facts of stock market
price statistics. This has recently been incorporated within
a macroscopic model of the stock market with a
competition between nonlinear trend-followers and nonlinear value investors
(Ide and Sornette (2001), Sornette and Ide (2001)).
We build on this insight and construct a very simple model of price dynamics,
which puts emphasis on the
fundamental {\it nonlinear} behavior of both classes of agents.

\smallskip
These well-known principles generate different kinds of risks between which
agents choose by arbitrage.
The former is a {\it competing risk}---Keynes (1936),
Orl\'ean (1989)---which leads
agents to imitate the collective point of view since the market price includes
it. Thus, it is assumed that Keynes' animal spirits may exist. More
simply, there is
the risk of mistaken expectation: agents believe in a price different from the
market price. Keynes uses
his famous beauty contest as a parable for stock markets. In order to predict
the winner of beauty contest, objective beauty is not very important, but knowledge
or prediction of others' prediction of beauty is. In
Keynes' view,
the optimal strategy is not to pick those faces the player thinks the prettiest,
but those the other players are likely to think the average opinion will be, or
those the other players will think the others will think the average 
opinion will be,
or even further along this iterative loop.

On the other hand, in the latter case, the emerging price is not necessarily in
harmony with economic reality and
fundamental value. Self-referred decisions and self-validation phenomena can
then indeed lead to speculative bubbles or sunspots---Azariadis (1981),
Azariadis-Guesnerie
  (1982),
Blanchard-Watson (1982),
Jevons (1871),
Kreps (1977). Thus, the latter risk is
the result of {\it precaution}. It addresses the fitting of market price to
fundamental value, and by extension, collapse of the speculative bubble.

Both attitudes are likely to be important and are integrated in
decision
rules. Agents realize an arbitrage between the two kinds of risk we have
described. That is why they
have both a mimetic behavior and an antagonistic one: they either
follow the collective point of view or they
have reversed expectations.

We are now going to put these assumptions into the simplest
possible mathematical form.
We assume that, at any given time $t$, the population is divided into two
parts.
Agents are explicitly
differentiated as being bullish or bearish in proportion $p_t$, and
$q_t=1-p_t$,
respectively.
The first ones expect an increase of the price, while the bearish ones expect a
decrease. The agents then form their opinion for time $t+1$ by sampling the
expectations of $m$ other
agents at time $t$, and modifying their own expectations accordingly.
The number $m$ of agents polled by a given agent to form her opinion at
time $t+1$ is the first important
parameter in our model.

We then introduce threshold densities $\rho_{hb}$ and
$\rho_{hh}$. We assume
$0\le\rho_{hb} \le \rho_{hh}\le1$. A bullish agent will change opinion
if at least one of the following
propositions is true:
\item{1.}At least $m\cdot\rho_{hb}$ among
the $m$ agents inspected are bearish.
\item{2.} At least $m\cdot\rho_{hh}$ among
the
$m$ agents inspected are bullish.

\smallskip
The first case corresponds to ``following the crowd,'' while the second case
corresponds to the ``antagonistic behavior.'' The quantity $\rho_{hb}$ is
thus the threshold for a bullish agent (``haussier'') to become bearish
(``baissier'') for mimetic reasons,
and similarly, $\rho_{hh}$ is the threshold for a bullish
agent to become bearish because there are ``too many'' bullish agents.
One reason for this behavior is, as we said, that the 
deviation of the market price from fundamental value is felt to be 
unsustainable.
Another reason is that if many managers tell you that they are bullish, it
is probable that they have large ``long'' positions in the market: they therefore
tell you to buy, hoping to be able to unfold in part their position in
favorable conditions with a good profit.

The deviation of the
threshold $\rho_{hb}$ above the symmetric value $1/2$
is a measure of the ``stubbornness'' (or ``buy-and-hold''
tendency) of the agent to keep her
position. For $\rho_{hb}=1/2$,
the agent strictly endorses without delay the opinion of the majority and believes
in any weak trend. This corresponds to a reversible dynamics.
A value $\rho_{hb}>1/2$ expresses a tendency towards conservatism:
a large $\rho_{hb}$ means that the agent will rarely
change opinion.
She is risk-adverse and would like to see an almost unanimity appearing before
changing her mind. Her future behavior has thus a strong memory of
her past position.
$\rho_{hb}-1/2$ can be called the bullish ``buy-and-hold'' index.

The deviation of the threshold $\rho_{hh}$ below $1$ quantifies the strength
of disbelief of the agent in the sustainability of a speculative trend.
For $\rho_{hh}=1$, she always follows the crowd and is never contrarian.
For $\rho_{hh}$ close to $1/2$, she has little faith in trend-following
strategies and is closer to a fundamentalist, expecting the price to revert
rapidly to its fundamental value. $1-\rho_{hh}$ can be called the
bullish reversal index.

Putting the above rules into mathematical equations we see that
the
probability $P$ for an agent who is bullish at time $t$ to change his
opinion at time
$t+1$ is:
\be
P \,=\, {\rm Prob}\left(\left \{ x < m\cdot(1-\rho_{hb})\right \}\cup
\left \{x> m\cdot \rho_{hh}\right \}\right )~,
\label{rules}
\ee
where $x$ is the number of bullish agents found in the sample of $m$ agents.

In an entirely similar way, we introduce thresholds $\rho_{bh}$, and
$\rho_{bb}$.
The thresholds  $\rho_{bh}$ and $\rho_{bb}$ have completely symmetric roles
when the agent is initially bearish.
$\rho_{bh}-1/2$ can be called the bearish ``buy-and-hold'' index.
$1-\rho_{bb}$ can be called the bearish reversal index.
The probability $Q$ for a bearish agent at time $t$ to become bullish
at time
$t+1$ is: $$ Q \,=\, {\rm Prob}\left(\left \{ x < m\cdot(1-\rho_{bh})\right
\}\cup \left \{x> m\cdot \rho_{bb}\right \}\right )~. $$

We can combine these two rules into a {\it dynamical law} governing the
time
evolution of the populations.
Denoting $p_{t}$ the proportion of bullish agents in the
population at time
$t$, we can find the new proportion, $p_{t+1}$, at time $t+1$, by taking into
account
those agents which have changed opinion according to the {\it deterministic
law}
given above.
To simplify notation, we let $p_{t+1}=p'$ and $p_{t}=p$. Then, the above
statements are easily used to express $p'$ in terms of $p$, by using the
probability of finding $j$ bullish people among $m$ (Corcos (1993)):
\bea
  p' \, &= & \, p-p\cdot\sum_{{j\geq m\cdot\rho_{hb}\atop {\rm or~} j <
m\cdot(1-\rho_{hh})} } {m\choose j} p^{m-j} (1-p)^j   \nonumber \\
&+&
(1-p)\cdot\sum_{{j\geq m\cdot\rho_{bh}\atop {\rm or~} j< 
m\cdot(1-\rho_{bb})}} {m
\choose j} (1-p)^{m-j} p^j \,   \label{anne} \\
&\equiv&  F_{\underline\rho,m}(p)~,  \nonumber
\eea
where
$\underline\rho=\{\rho_{hb},~\rho_{bh},~\rho_{hh},~\rho_{bb}\}$.
Thus, the function $F_{\underline\rho,m}(p)$ completely characterizes
the dynamics of the proportion of bullish and bearish populations.

\section{Qualitative analysis of the dynamical properties}

\subsection{The limit $m \to \infty$}

\begin{figure}
\begin{center}
\includegraphics[width=15cm]{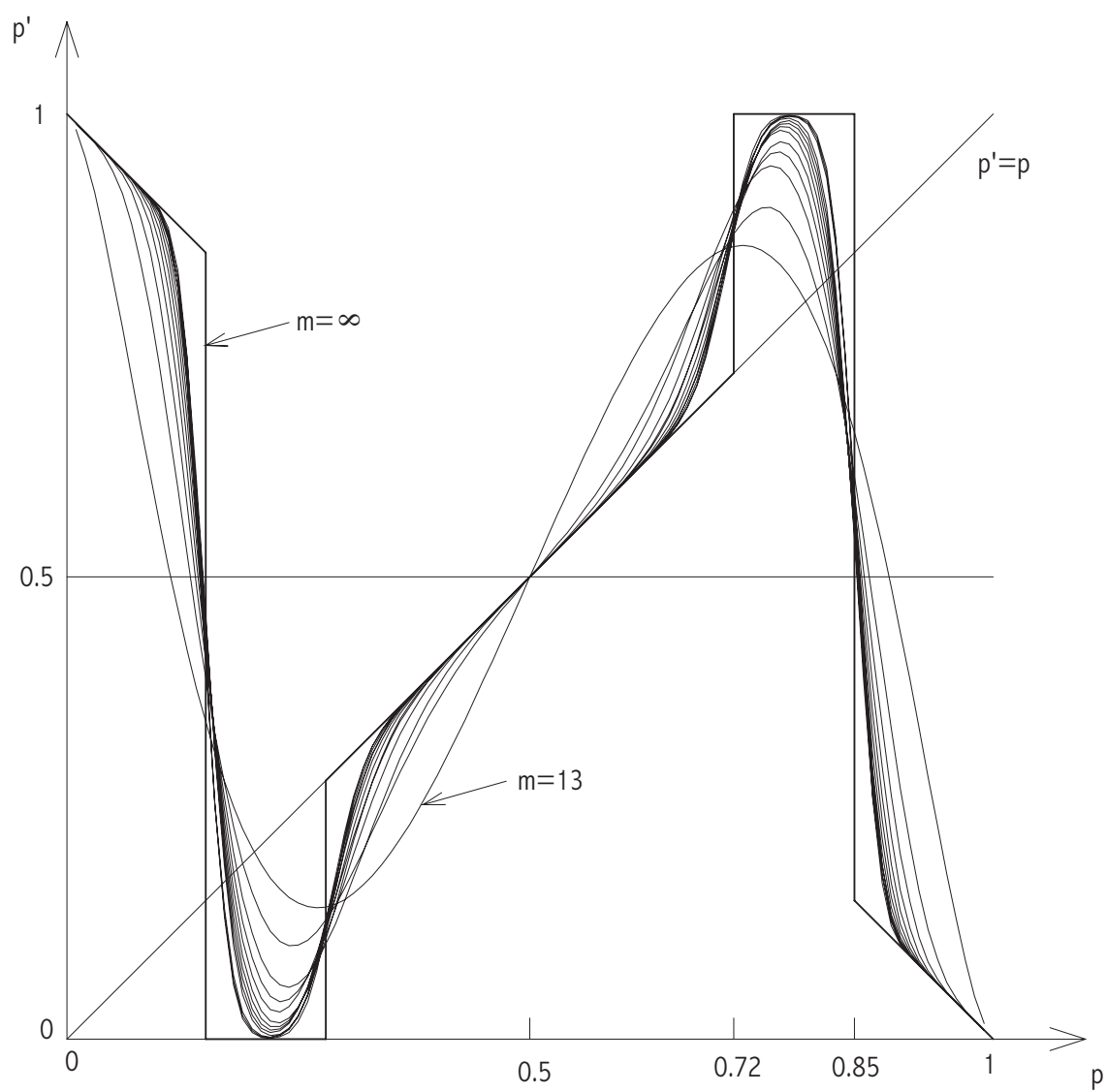}
\caption{\label{fig1} The family of functions $F_{\underline\rho,m}(p)$
for $\rho_{hb} = \rho_{bh} = 0.72$ and $\rho_{hh}=\rho_{bb} = 0.85$. 
The curves are for $m = 13 + j \cdot 26, j = 0, \dots, 13$. 
Note the convergence to the function $G_{\underline\rho}$, (indicated by $m = \infty$).}
\end{center}
\end{figure}

The law given by Eq.\ref{anne} is not easy to analyze, and we give in 
Fig.~\ref{fig1}
a few sample curves $F_{\underline\rho,m}$.
We see that as $m$ gets larger, the curves seem to tend to a limiting curve.
Using this observation,
our conceptual understanding of the dynamics can be drastically
simplified
if we consider the problem for a large number $m$ of polled partners.
Indeed, it is most convenient to first study the unrealistic problem
$m=\infty$ and to view the large $m$ case as a perturbation of this limiting
case.
The main ingredient in the study of the case $m=\infty$ is the Law of Large
Numbers, which we use in a form given in Feller (1966):\\
{\bf Lemma}. Let $g$ be a continuous function on $[0,1]$. Then, for
$s\in[0,1]$,
\be
\lim_{m\to\infty }\sum_{j=0}^m{m\choose j} s^j (1-s)^{m-j}\cdot g(j/m) \,=\,
g(s)~.
\ee

We apply this lemma to the (piecewise continuous) function
$g=f_h$, where $f_h$ is the indicator function of the
set defining $P$:
\be
f_h \,=\, \cases { 1~, & if $x\ge\rho_{hb}$ or $x < 1-\rho_{hh}$,\cr
0~, & otherwise~.\cr }
\ee
Similarly, we define
\be
f_b \,=\, \cases { 1~, & if $x\ge\rho_{bh}$ or $x < 1-\rho_{bb}$,\cr 
0~, & otherwise~.\cr }
\ee
It is now easy to check that the lemma implies
\be
\lim_{m\to\infty} F_{\underline\rho,m}(p)
\,=\, p-p\cdot f_h(1-p)+(1-p)\cdot f_b(p)\,\equiv\,G_{\underline\rho}(p)~.
\label{limit}
\ee
Note again that we do not consider $G_{\underline\rho}(p) $ itself as an
evolution law for the population of bullish agents, but 
$G_{\underline\rho}$ can serve
very well as an
approximation for the true laws $F_{\underline\rho,m}$ for large $m$. 
In Fig.~\ref{fig1}
we show how the functions $F_{\underline\rho,m}$ converge to
$G_{\underline\rho}$.

\subsection{Classification of the different regimes}

\begin{figure}
\begin{center}
\includegraphics[width=15cm]{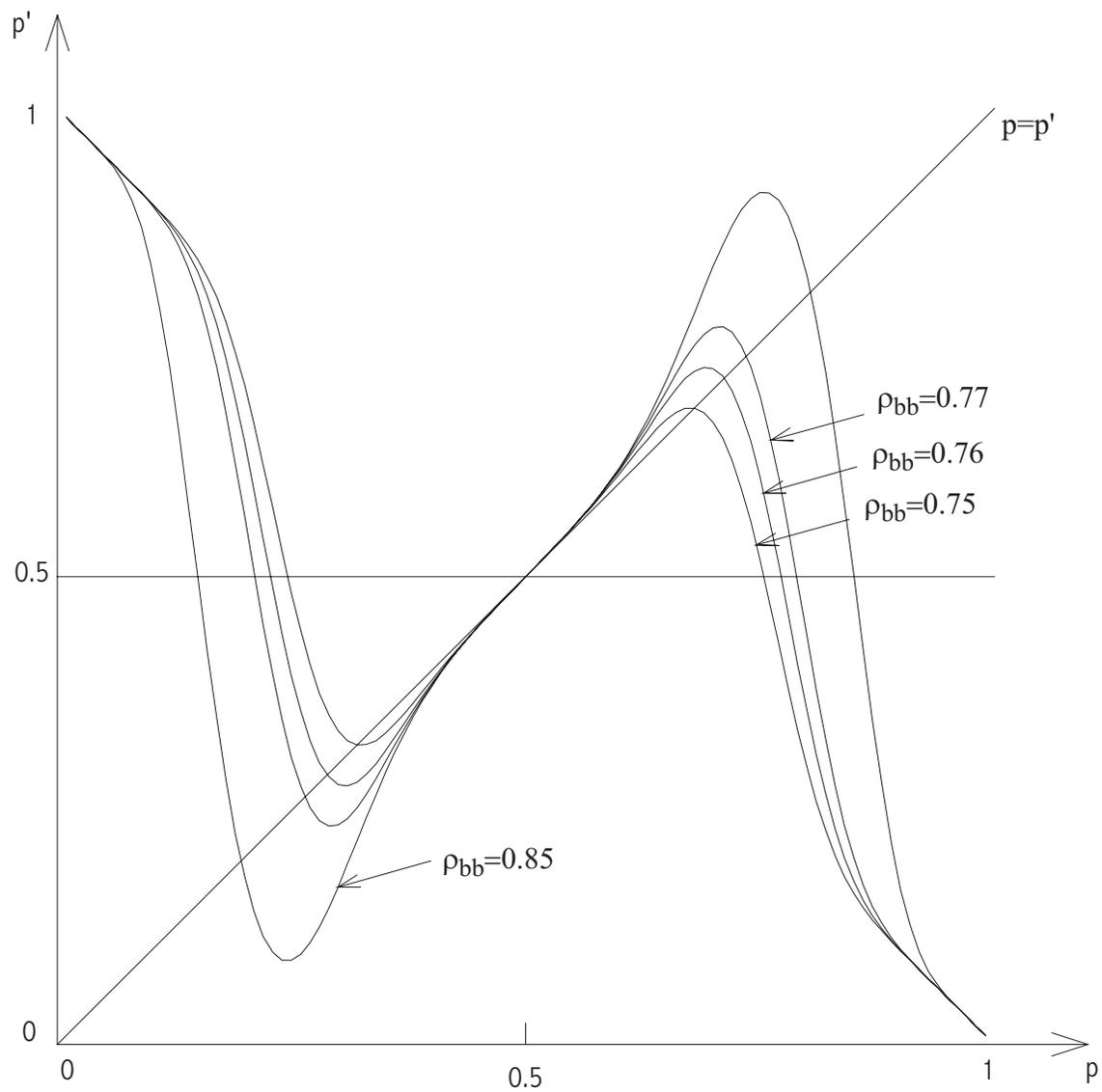}
\caption{\label{fig2} Four curves $F_{\underline\rho,m}$, for $m=60$ and
$\rho_{hb}=\rho_{bh}=0.72$, with $\rho_{hh}=\rho_{bb}=0.75, 0.76, 0.77, 0.85$.}
\end{center}
\end{figure}

\begin{figure}
\begin{center}
\includegraphics[width=15cm]{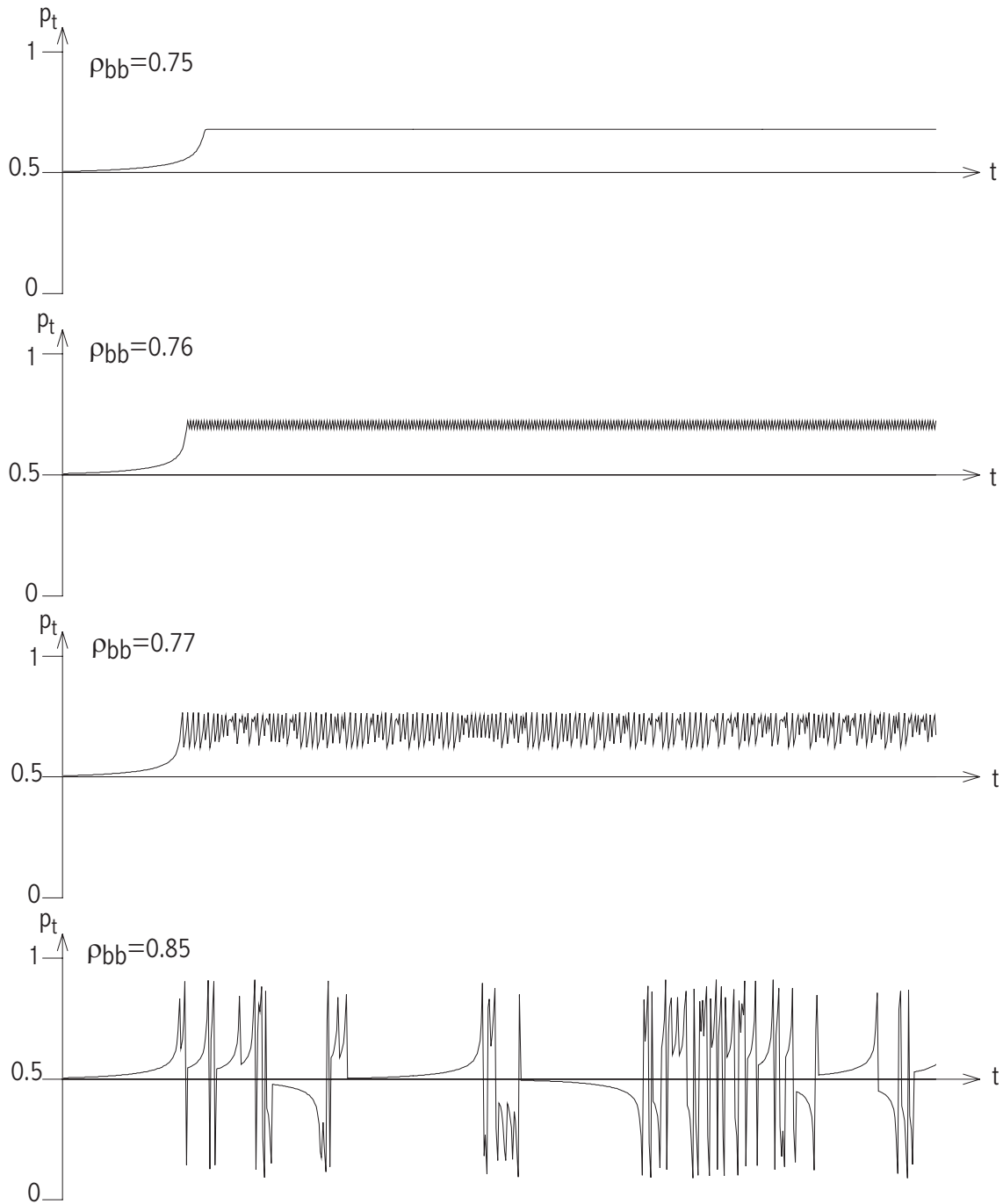}
\caption{\label{fig3} The time series for the same parameter values
as in Fig.~2. Note that, for
$\rho_{hh}=\rho_{bb}=0.75$, one has convergence to a bullish equilibrium, for
$0.76$ a bullish period $2$, for $0.77$ a bullish, but chaotic behavior. The
most interesting case is $\rho_{hh}=\rho_{bb}=0.85$, where calm periods
alternate in a seemingly random fashion with speculative bubbles.}
\end{center}
\end{figure}

In the preceding section, we have shown how to gain a qualitative understanding
of the maps $F_{\underline\rho,m}$, when $m$ is large.
We can now apply in a rather straightforward way the
general theory of 1-dimensional discrete time dynamical systems (see {\it e.g.},
Collet-Eckmann (1980)) to the functions $F_{\underline\rho,m}$.
The recurrence $p_t \rightarrow p_{t+1}$ can exhibit several typical behaviors
which, for large $m$ depend essentially only on the
set of parameters ${\underline\rho}$. We enumerate a few of
them and refer the reader to Figs.~\ref{fig2} and \ref{fig3}. In this 
section, we restrict
our attention to the symmetric case $\rho_{hb}=\rho_{bh}$ and
$\rho_{hh}=\rho_{bb}$.

\item{1.}The most trivial case is the appearance of a stable fixed point.
This will occur when the buy-and-hold index $\rho_{hb}-1/2$ is not
too large and the reversal index $1-\rho_{hh}$ is not too small.
For example, this occurs for $\rho_{hh}=\rho_{bb}=0.75$,
$\rho_{hb}=\rho_{bh}=0.72$, and $m=60$. Then, the population
will equilibrate, and converge to
$p=1-q\approx0.68$, or to $p\approx0.32$ (see upper panel of figure 
\ref{fig3}).

\item{2.} The next more interesting case is the appearance of a limit cycle
(of period 2): at successive times, the population of bullish and bearish agents
will oscillate between two different values. This happens, {\it e.g.}, for
$\rho_{hh}=\rho_{bb}=0.76$, with the other parameters as before
(see second panel of figure \ref{fig3}).

\item{3.} But for certain values of the parameters, {\it e.g.},
$\rho_{hh}=\rho_{bb}=0.85$, the
sequence of values of $p_t$ is a {\it chaotic} sequence, with positive Liapunov
exponent (cf.~Eckmann-Ruelle (1985)). The mechanism for this is really a
combination of sufficiently strong buy-and-hold index $\rho_{hb}-1/2$ and of
sufficiently weak reversal index $1-\rho_{hh}$.
This regime thus occurs when the opinion of a trader has a strong memory of
her past positions and changes it only when a strong majority appears. This
regime also requires a weak belief of the agent in fundamental valuation, as
she will believe until very late that a strong bullish or bearish 
speculative trend
is sustainable. Fundamentally,
it is this {\it self-referential} behavior of the anticipations {\it alone}
which is responsible for a deterministic, but seemingly erratic evolution of
the population of bullish and bearish agents.
No external noise is needed to make this happen,
and in general, we view external stimuli as acting on top of the intrinsic
mechanism which we exhibit here (Eckmann (1981)). Note that the set 
of parameter values
${\underline \rho}$ for
which chaos is expected (say, near the values used at the bottom of
Fig.~\ref{fig3}) has positive Lebesgue measure.
\vskip\baselineskip

We next consider in more detail the time evolution of $p_t$
for the parameter values of the last frame of Fig.~\ref{fig3},
which are typical for the abundant set of ``chaotic'' parameter values, and we
will show how the time evolution exhibits
``speculative bubbles.'' This phenomenon
is akin to the notion of intermittency (of ``Type I'')
as known to physicists, see
{\it e.g.}, Manneville (1991)
for an exposition. Indeed, we can distinguish two
distinct behaviors in the last frame of Fig.~\ref{fig3}, which occur 
repeatedly with more
or less
pronounced separation. The first process is the ``laminar phase,'' 
which is seen
to occur when the population $p_t$ is near 0.5. Then, the evolution of the
population is slow, and the population grows slowly away from 0.5, either
monotonically or through an oscillation of period 2, depending on
${\underline
\rho}$. This motion is slower when the inspected sample size ($m$) is larger,
reflecting a more stable evolution for less independent agents. When the
distance from 0.5 is large, erratic behavior sets in, which persists until the
population reaches again a value of about 0.5, at which point the 
whole scenario
repeats. The determinism of the model is reflected
by ``equal causes lead to equal effects,'' while its chaotic nature 
is reflected
by the erratic length of the laminar periods, as well as of the bubbles of wild
behavior.

Having analyzed qualitatively the evolution of the number of bullish agents,
we next describe how the price  $\pi_{t+1}$ of
an asset at time $t+1$ is related to
the proportion $p_t$ of bullish agents. One can argue (Corcos (1993), 
Bouchaud and Cont (1998), Farmer (1998))
that the price change $\pi_{t+1}-\pi_t$ from one period to the next is a {\it 
monotone} function of $p_t$ (and, perhaps,  of $\pi_t$). This function 
is positive when $p_t>1/2$ and negative when $p_t<1/2$. If the reaction to 
a change in $p_t$ is reflected in the prices in the next period, then a bubble 
in $p_t$ will lead to a speculative bubble in the prices in the next period.
Thus, our model
predicts the occurrence of bubbles from the behavior of the agents alone.
Furthermore, for quite general laws of the form
\be
\pi_{t+1}\,=\,H(\pi_t,p_t)~,
\ee
a simple application of the chain rule of differentiation leads to the
observation that the variable $\pi_t$ has the same Liapunov exponent as $p_t$.
In fact, this will be the case if $0<\partial_\pi H
<\lambda$ and $\partial_p H>c>0$, where $
\lambda $ is the Liapunov exponent for $p_t$,
as follows from
$\delta \pi_{t+1} = \partial_{\pi} H \cdot\delta \pi_t + \partial_p H 
\cdot\delta p_t$.
This condition is, in particular, satisfied for a law of the form
$\pi_{t+1}=\pi_t+G(p_t)$,
where $G$ is strictly monotone.
Thus, chaotic behavior of bullish agents leads to chaotic behavior of prices.

In the sequel, we shall take the simplest form of a log-difference of the
price linearly proportional to the order unbalance (Farmer (1998)), leading to
\be
\ln \pi_{t+1} - \ln \pi_{t+1} \equiv r_{t+1} = \gamma (p_t - {\textstyle{1 \over 2}})~,
\label{retuneq}
\ee
showing that the return $r_t$ calculated over one period is proportional to
the imbalance $p_t - {1 \over 2}$. Thus, the properties of the return time
series can be derived directly from those of $p_t$ as we document below.

To summarize this qualitative analysis of the case of an infinite number $N$
of agents, we observe a time evolution which,
while satisfying certain criteria of randomness (such as possessing an
absolutely continuous invariant measure and exhibiting a positive Liapunov
exponent---cf.~Eckmann-Ruelle (1985)) at the same time
exhibits some regularities on short time scales, since it is deterministic.
Our model thus establishes that straightforward fundamental conditions may
suffice to generate chaotic stock market behavior, depending on the 
parameter values.
If the market adjusts
present market price on the basis of expectations and mimicry---self-referred
behavior---then chaotic evolution of the population will also imply chaotic
evolution of prices.

\section{Quantitative analysis of the speculative bubbles within the 
chaotic regime
in the symmetric case}

For an infinite number $N$ of agents and
in the symmetric case $\rho_{hb}=\rho_{bh}\equiv \rho_1$ and
$\rho_{hh}=\rho_{bb}\equiv \rho_2$,
let us rewrite the dynamical evolution (\ref{anne}) of the system as
\be
p'=p-p \sum_{j=0}^m {m \choose j} p^{m-j} (1-p)^j f \left( 
\frac{j}{m} \right)+ (1-p)
\sum_{j=0}^m {m \choose j} (1-p)^{m-j} p^j f \left( \frac{j}{m} \right),
\ee
where
\be
f(x)=\left\{ \begin{array}{l}
1,~ \mbox{if}~  x \ge \rho_1~ \mbox{or}~  x < 1- \rho_2 \\
0,~ \mbox{otherwise}.
\end{array}
\right.
\ee
Let us define
\be
g_m(p)=\sum_{j=0}^m {m \choose j} p^{m-j} (1-p)^j f \left( 
\frac{j}{m} \right)~,
\ee
which yields
\be
p'= F_m(p) = p-p \cdot g_m(p) + (1-p) \cdot g_m(1-p)~.
\label{general}
\ee
This expression (\ref{general}) generalizes (\ref{limit}) to arbitrary $m$.

\begin{figure}
\begin{center}
\includegraphics[width=15cm]{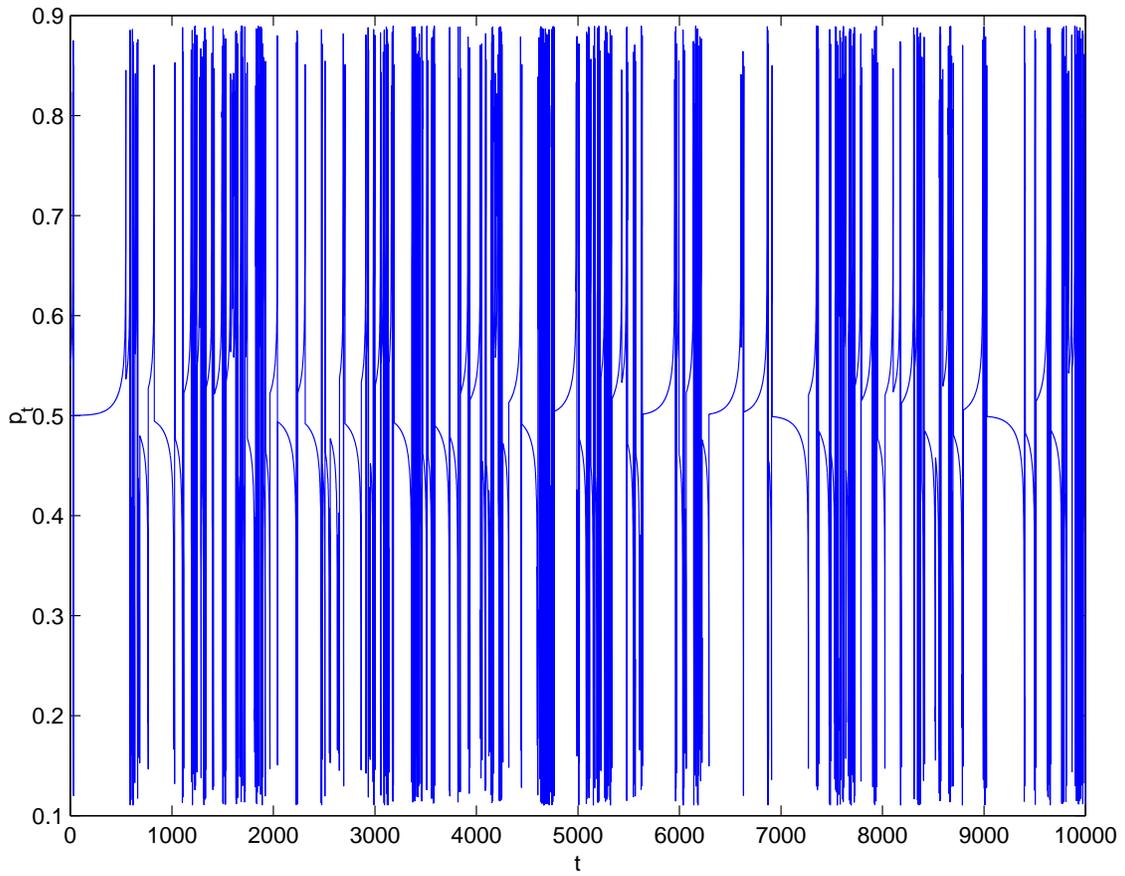}
\caption{\label{fig:1} Evolution of the system over $10000$ time steps
for $N=\infty$, $m=60$ polled agents and the parameters
$\rho_{hb}=\rho_{bh}=0.72$ and  $\rho_{hh}=\rho_{bb}=0.85$.}
\end{center}
\end{figure}

As was described in the previous section,
this system can exhibit chaotic behavior for certain values of the
parameters. An example is given in figure \ref{fig:1} which shows a long
time series, showing many positive bubbles and negative bubbles interrupted by
chaotic oscillatory phases. For the time being, we do not worry about the
existence of the negative bubbles, which are rarely if ever observed 
in real markets:
this is an artifact of the symmetry $\rho_{hb}=\rho_{bh}\equiv \rho_1$ and
$\rho_{hh}=\rho_{bb}\equiv \rho_2$, that we shall relax later. Keeping the
symmetry assumption simplified the theoretical analysis without changing
the key results obtained below.

\begin{figure}
\begin{center}
\includegraphics[width=15cm]{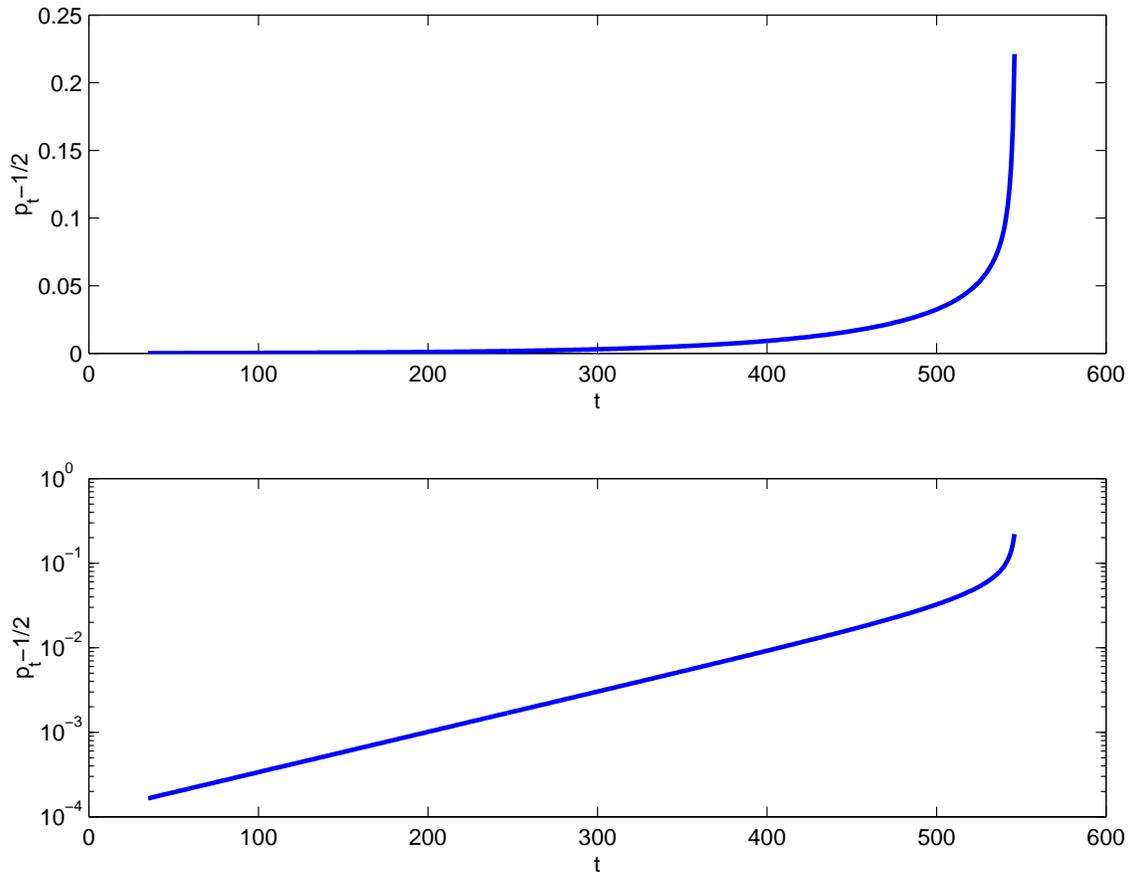}
\caption{\label{fig:2} The first bubble of figure \ref{fig:1}
for $N=\infty$ agents with $m=60$ polled agents and
parameters $\rho_{hb}=\rho_{bh}=0.72$ and $\rho_{hh}=\rho_{bb}=0.85$.}
\end{center}
\end{figure}

Let us consider the first bubble developing in the
time interval from $t=35$ to $t=546$ as seen in figure \ref{fig:2}-a).
Figure \ref{fig:2}-b plots the logarithm of $p-1/2$ as a function of
linear time: the linear trend from $t=35$ to $t \approx 480$ seen
in the lower panel qualifies
an exponential growth $p-1/2 \propto e^{\kappa t}$ (with $\kappa >0$) 
followed by
a super-exponential growth accelerating so much as to give the impression
of reaching a singularity in finite-time.

\begin{figure}
\begin{center}
\includegraphics[width=15cm]{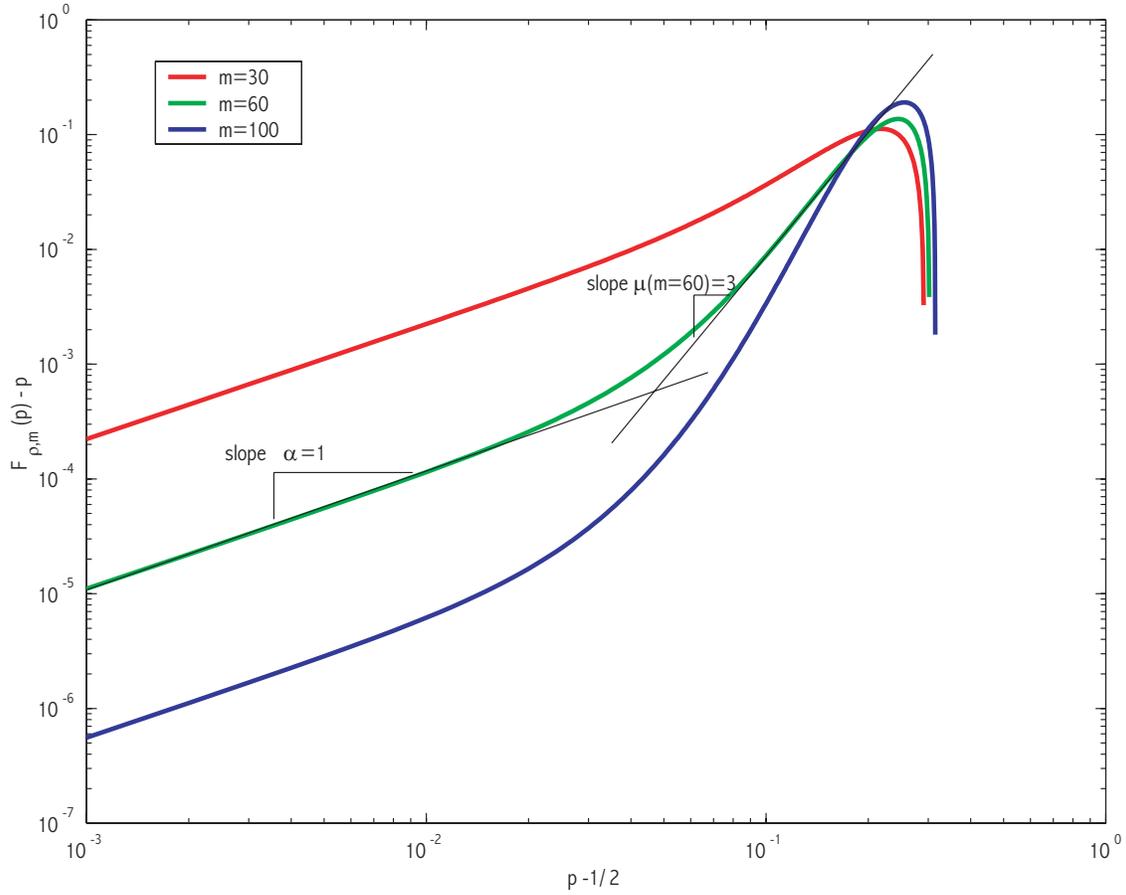}
\caption{\label{fig:3} The logarithm of
$F_m(p)-p$ versus the logarithm of $p-1/2$ for
three different values of $m=30, 60$ and $100$,
with $\rho_{hb}=\rho_{bh}=0.72$ and $\rho_{hh}=\rho_{bb}=0.85$.
Note the transition from a slope $1$ to a large effective slope
before the reinjection due to the contrarian mechanism.
}
\end{center}
\end{figure}

To understand this phenomenon, we plot the logarithm of
$F_m(p)-p$ versus the logarithm of $p-1/2$ in figure \ref{fig:3} for
three different values of $m=30, 60$ and $100$.
Two regimes can be observed.
\begin{enumerate}
\item For small $p-1/2$, the slope of $\log_{10} (F_m(p)-p)$ versus 
$\log_{10} (p-1/2)$ is $1$, {\it i.e}, 
\be
p'-p \equiv F_m(p)-p \simeq \alpha(m) \left(p-\frac{1}{2}\right)~.
\label{linearexp}
\ee
This expression (\ref{linearexp}) explains the exponential growth
observed at early times in figure \ref{fig:2}.

\item For larger $p-1/2$, the slope of $\log_{10} (F_m(p)-p)$ versus
$\log_{10} (p-1/2)$ increases above $1$ and stabilizes to a value ${\mu(m)}$
before decreasing again due to the reinjection produced by the contrarian
mechanism. The interval in $p-1/2$ in which the slope is approximately
stabilized at the value ${\mu(m)}$ enables us to write
\be
F_m(p)-p \simeq \beta(m) \left(p-\frac{1}{2}\right)^{\mu(m)}~,~~~~~
\mbox{with}~ \mu>1~.
\ee
\end{enumerate}

These two regimes can be summarized in the following phenomenological
expression for $F_m(p)$:
\bea
F_m(p)&=&\frac{1}{2} + \left(1 - 2 g_m(1/2)  - g_m'(1/2) \right)
\left(p-\frac{1}{2} \right) + \beta(m) \left(p-\frac{1}{2} 
\right)^{\mu(m)}~, \\
&=& \frac{1}{2} + \left(p-\frac{1}{2} \right) +\alpha(m) \left(p-\frac{1}{2}
\right) + \beta(m) \left(p-\frac{1}{2} \right)^{\mu(m)}~~~~\mbox{with}~ \mu>1~,
\label{hypcheck}
\eea
and
\be
\alpha(m)=  -2 g_m(1/2)  - g_m'(1/2)~.
\ee
This expression can be obtained as an approximation of the
exact expansion derived in the Appendix.

\begin{figure}
\begin{center}
\includegraphics[width=15cm]{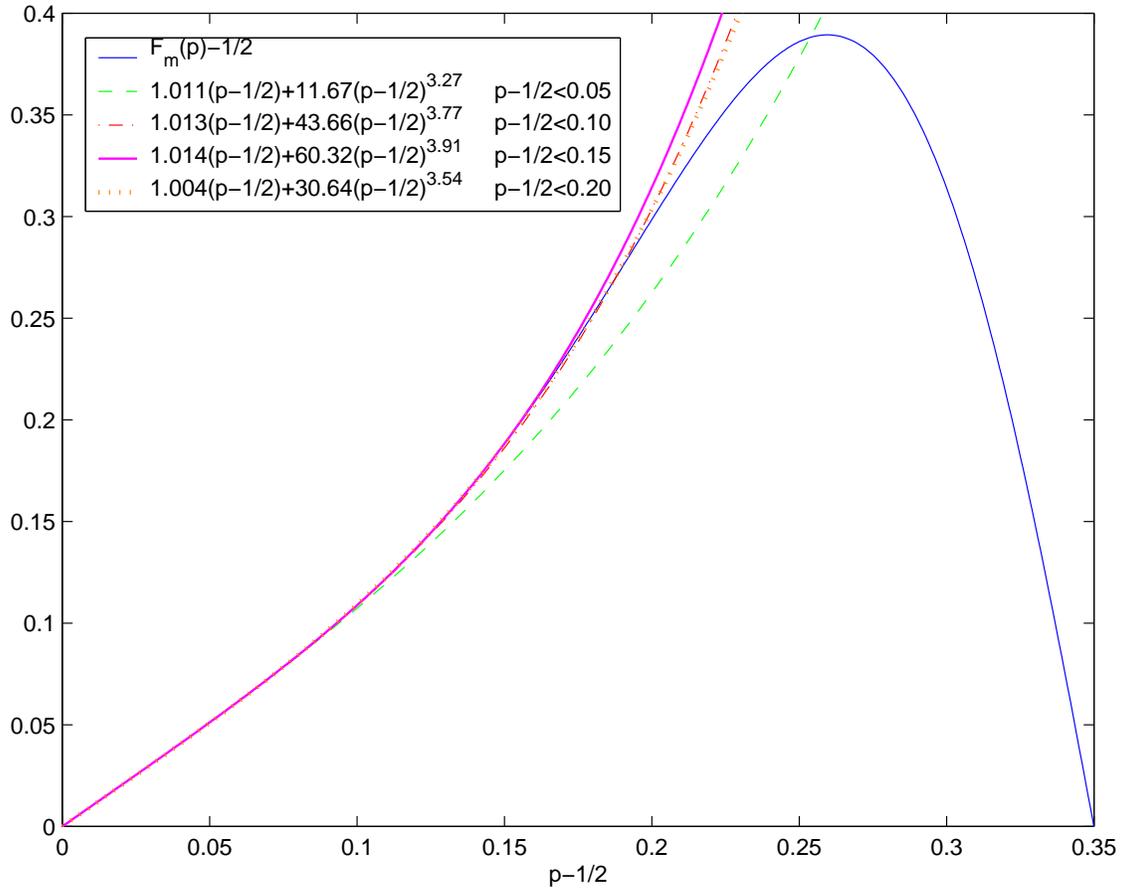}
\caption{\label{fig:fit} Approximation of the function $F_m(p) -\frac{1}{2}$ by
the function $f(p)=[1+\alpha] \left( p+ \frac{1}{2} \right) + \beta  \left( p+
\frac{1}{2} \right)^\mu$ over different $p$-intervals, for $m=60$ 
interacting agents
and parameters  $\rho_{hb}=\rho_{bh}=0.72$ and $\rho_{hh}=\rho_{bb}=0.85$.}
\end{center}
\end{figure}

In order to check the hypothesis (\ref{hypcheck}), we numerically solve the
following problem
\be
\min_{\{\alpha, \beta, \mu\}} \left| \left| F_m(p)-\frac{1}{2} - [1+\alpha]
\left( p - \frac{1}{2} \right) - \beta  \left( p - \frac{1}{2} \right)^\mu
\right| \right|^2~,  \label{nhgnnlwla}
\ee
which amounts to constructing the best approximation of the exact map $F_m(p)$
in terms of an effective power law acceleration (see (\ref{nhgnald}) below).
The results obtained for $m=60$ interacting agents and
$\rho_{hb}=\rho_{bh}=0.72$ and $\rho_{hh}=\rho_{bb}=0.85$ are given in
table \ref{tab:1} and shown in figure \ref{fig:fit}.
The numerical values of $\alpha$ are in good agreement with the
theoretical prediction~: $ \alpha(m)=F_m'(1/2)-1$ which yields 
$\alpha(m) \simeq
0.011$ in the present case ($m=60$, $\rho_{hb}=\rho_{bh}=0.72$ and
$\rho_{hh}=\rho_{bb}=0.85$).
As a first approximation, we can consider that the exponent $\mu$ is fixed
over the interval of interest, which is reasonable according to the
very good quality of the fits shown in figure \ref{fig:fit}. We can conclude
from this numerical investigation that $\mu(m) \in [3,4]$. A finer
analysis shows however that the exponent $\mu$ is in fact not perfectly
constant but shifts slowly from about $3$ to $4$ as $p$ increases. This should
be expected as the function $F_m(p)$ contains many higher-order terms.
We can also note that the parameter
$p_c=\left(\beta / \alpha\right)^{-1/\mu}$, which defines
the typical scale of the crossover remains constant and
equal to $p_c \simeq 0.70$ for all the fits (except for the largest interval
$p-1/2  < 0.2$, for which  $p_c$ = 0.8). In sum, the procedure 
(\ref{nhgnnlwla})
and its results show that the effective power law representation
(\ref{hypcheck}) is a cross-over phenomenon: it is not dominated by the
``critical'' value $\rho_{hb}= \rho_{bh}$ of the jump of the map obtained
in the limit of large $m$.

\begin{table}[b]
\begin{center}
\begin{tabular}{|c||c c c|}
\hline
Optimization Domain & $\alpha$ & $\beta$ & $\mu$\\
\hline
$0 \le p-\frac{1}{2} \le 0.05$ & 0.011 & 11.67 & 3.27\\
$0 \le p-\frac{1}{2} \le 0.10$ & 0.013 & 43.66 & 3.77\\
$0 \le p-\frac{1}{2} \le 0.15$ & 0.014 & 60.32 & 3.91\\
$0 \le p-\frac{1}{2} \le 0.20$ & 0.004 & 30.64 & 3.54\\
\hline
\end{tabular}
\end{center}
\caption{\label{tab:1} Optimized parameters $\alpha$, $\beta$ and $\mu$ for
several optimization interval with m=60 interacting agents and
$\rho_{hb}=\rho_{bh}=0.72$ and $\rho_{hh}=\rho_{bb}=0.85$}
\end{table}

Introducing the notation $\epsilon=p-1/2$, the dynamics associated 
with the effective
map (\ref{hypcheck}) can be rewritten
\be
\epsilon'-\epsilon = \alpha(m) \epsilon + \beta(m) \epsilon^{\mu(m)},
\ee
which, in the continuous time limit, yields
\be
\frac{d \epsilon}{dt}= \alpha(m) \epsilon + \beta(m) \epsilon^{\mu(m)}~.
\label{nhgnald}
\ee
Thus, for small $\epsilon$, we obtain an exponential growth rate
\be
\epsilon_t \sim e^{\alpha(m)t}~,
\ee
while for large enough $\epsilon$
\be
\epsilon_t \sim (t_c-t)^{-\frac{1}{\mu(m)-1}}~.
\label{finitetimesingm}
\ee

\begin{figure}
\begin{center}
\includegraphics[width=15cm]{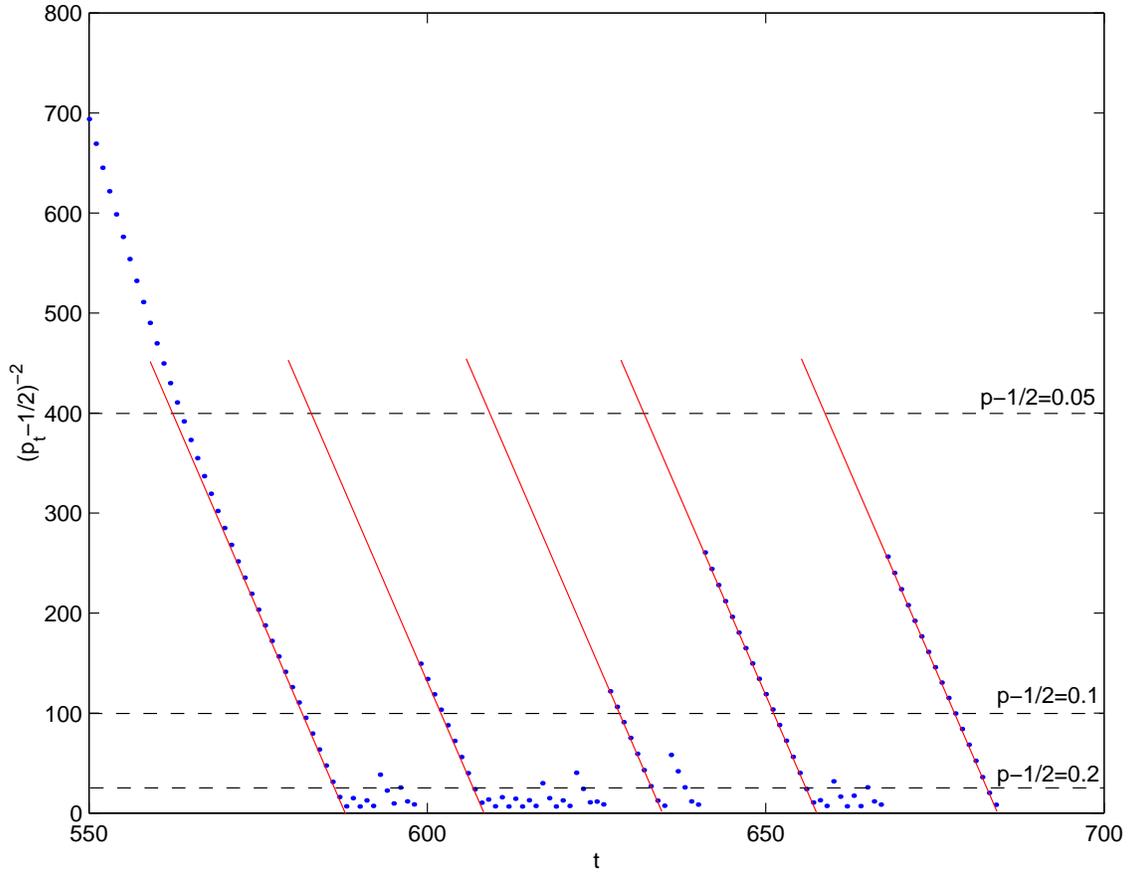}
\caption{\label{fig:4} $\frac{1}{(p_t-1/2)^2}$ versus $t$ to qualify the finite
time singularity predicted by (\ref{finitetimesingm60}) for
$m=60$ with $\rho_{hb}=\rho_{bh}=0.72$ and 
$\rho_{hh}=\rho_{bb}=0.85$. The points
are obtained from the time series $p_t$ and the straight continuous lines
are the best linear fits. The horizontal dashed lines indicate the levels
$p-1/2= 0.05, 0.01$ and $0.2$ to demonstrate that most of the visited values
are close to the unstable fixed point, which determines the effective value
of the nonlinear exponent $\mu \approx 3$.
}
  \end{center}
\end{figure}

For example, for $m=60$ with $\rho_{hb}=\rho_{bh}=0.72$ and
$\rho_{hh}=\rho_{bb}=0.85$, we can check on figure \ref{fig:3} that
$\mu(m)=3$, which yields for large $\epsilon$:
\be
p_t-\frac{1}{2} \sim \frac{1}{\sqrt{t_c-t}}~.
\label{finitetimesingm60}
\ee
The prediction (\ref{finitetimesingm60}) implies that
plotting $(p_t-1/2)^{-2}$ as a function of $t$ should be a straight 
line in this
regime. This non-parametric test
is checked in figure \ref{fig:4} on five successive bubbles. This provides
a confirmation of the effective power law representation 
(\ref{hypcheck}) of the
map. The fact that it is the lowest estimate $\mu \approx 3$ shown
in table \ref{tab:1} which dominates
in figure \ref{fig:4} results
simply from the fact that it is the longest transient
corresponding to the regime where $p$ is closest to
the unstable fixed point $1/2$. This is visualized in figure \ref{fig:4}
by the horizontal dashed lines indicating the levels
$p-1/2= 0.05, 0.01$ and $0.2$. This demonstrates that most of the 
visited values
are close to the unstable fixed point, which determines the effective value
of the nonlinear exponent $\mu \approx 3$.

With the price dynamics (\ref{retuneq}), the prediction (\ref{finitetimesingm})
implies that the returns $r_t$ should increase in an accelerating
super-exponential fashion
at the end of a bubble, leading to a price trajectory
\be
\pi_t = \pi_c - C (t_c-t)^{\frac{\mu(m)-2}{\mu(m)-1}}~,
\label{pricetrajbubble}
\ee
where $\pi_c$ is the culminating price of the bubble reached at $t=t_c$
when $\mu(m) >2$, such the finite-time singularity in $r_t$ gives rise
only to an infinite slope of the price trajectory. The behavior
(\ref{pricetrajbubble}) with an exponent $0 < \frac{\mu(m)-2}{\mu(m)-1} < 1$
has been documented in many bubbles (Sornette {\it et al.} (1996), Johansen {\it et al.} (1999,
2000), Johansen and Sornette (1999, 2000), Sornette and Johansen (2001), Sornette
and Andersen (2001), Sornette (2001)).
The case $m=60$ with $\rho_{hb}=\rho_{bh}=0.72$ and
$\rho_{hh}=\rho_{bb}=0.85$ shown in figure \ref{fig:3} leads to
$\frac{\mu(m)-2}{\mu(m)-1}=1/2$, which is in reasonable agreement with 
previously reported values.

Interpreted within the present model,
the exponent $\frac{\mu(m)-2}{\mu(m)-1}$ of the price singularity gives
an estimation of the ``connectivity'' number $m$ through the dependence
of $\mu$ on $m$ documented in figure \ref{fig:3}. Such a relationship
has already been argued by Johansen {\it et al.}, (2000) at a phenomenological level using 
a mean-field equation in which the exponent is directly related to the number
of connections to a given agent.

\section{Statistical properties of price returns in the symmetric case}

\begin{figure}
\begin{center}
\includegraphics[width=15cm]{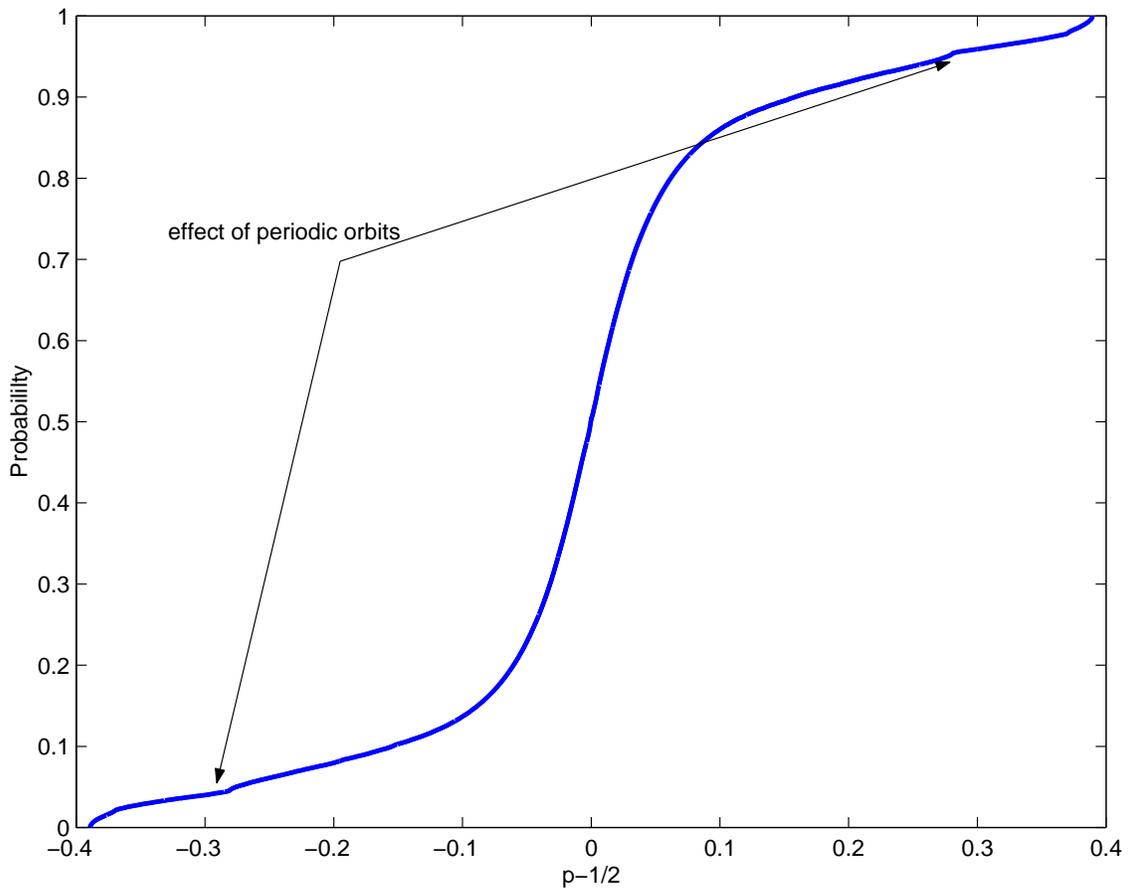}
\caption{\label{fig:5} Cumulative distribution for $m=60$ polled agents and the
parameters $\rho_{hb}=\rho_{bh}=0.72$ and $\rho_{hh}=\rho_{bb}=0.85$.}
\end{center}
\end{figure}

Using the price dynamics (\ref{retuneq}), the distribution of
$p-1/2$ is the same as the distribution of returns, which is the first
statistical property analyzed in econometric work 
(Campbell {\it et al.} (1997), Lo and MacKinlay (1999), Lux (1996), Pagan (1996), 
Plerou et al (1999), Laherr\`ere and Sornette (1998)).
Note that the distribution of $p-1/2$ is nothing but the invariant
measure of the chaotic map $p'(p)$ which can be shown to be continuous
with respect to the Lebesgue measure (Eckmann and Ruelle (1985)).
Figure \ref{fig:5} shows the cumulative distribution of $r_t \propto p_t-1/2$.
Notice the two breaks at $|p-1/2|=0.28$, which are due
to the existence of weakly unstable periodic orbits corresponding to 
a transient oscillation between bullish and bearish states.

\begin{figure}
\begin{center}
\includegraphics[width=15cm]{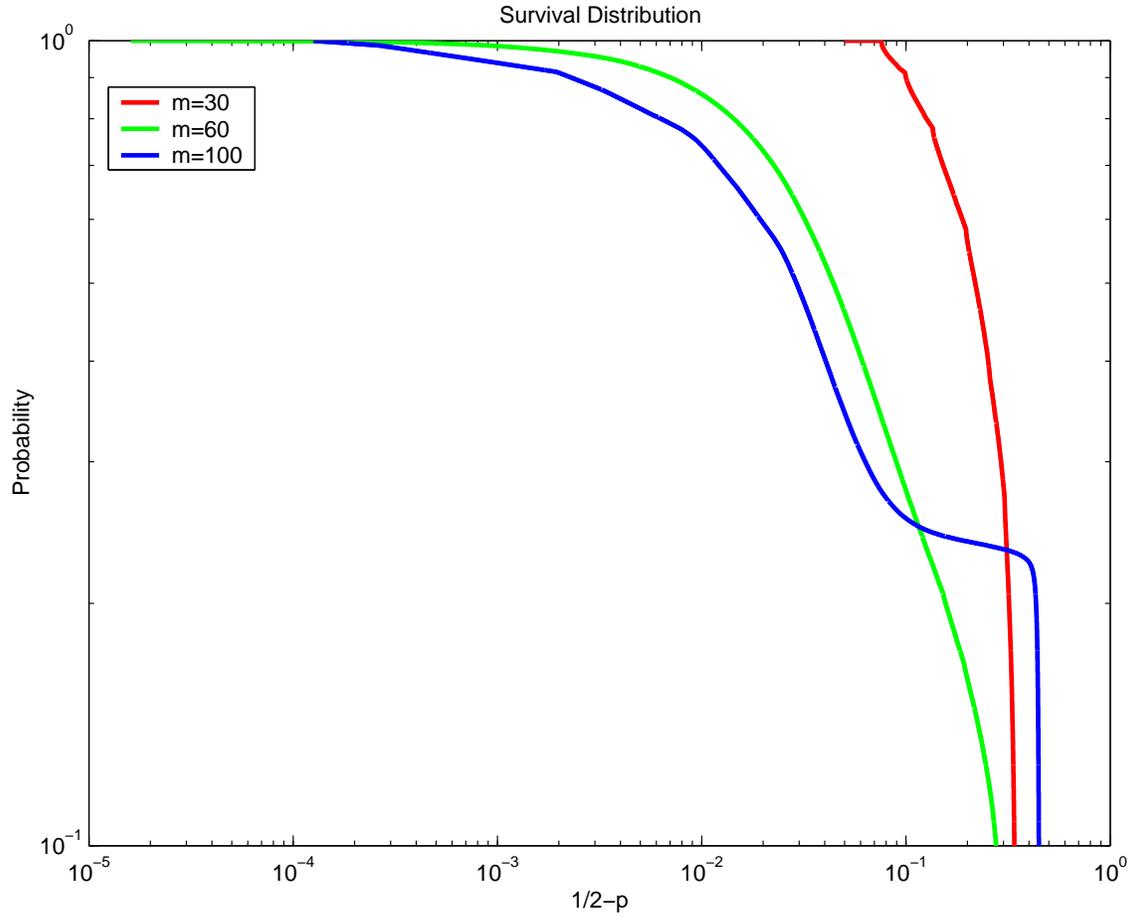}
\caption{\label{fig:6} Survival distribution for $m=30, 60$ and $100$ 
polled agents and
parameters $\rho_{hb}=\rho_{bh}=0.72$ and  $\rho_{hh}=\rho_{bb}=0.85$}
\end{center}
\end{figure}

Figure \ref{fig:6} plots in double logarithmic scales
the survival distribution of $r_t \propto p_t-1/2$
for $m=30, 60$ and $100$. For $m=60$, we can observe an
approximate power law tail but the exponent is smaller than $1$
in contradiction with the empirical evidence which suggests
a tail of the survival probability with exponents $3-5$. In the other
cases, we cannot conclude on the existence of a power law regime, but it is
obvious that the tail behavior of the distribution function depends on
the number $m$ of polled agents.

\begin{figure}
\begin{center}
\includegraphics[width=8cm]{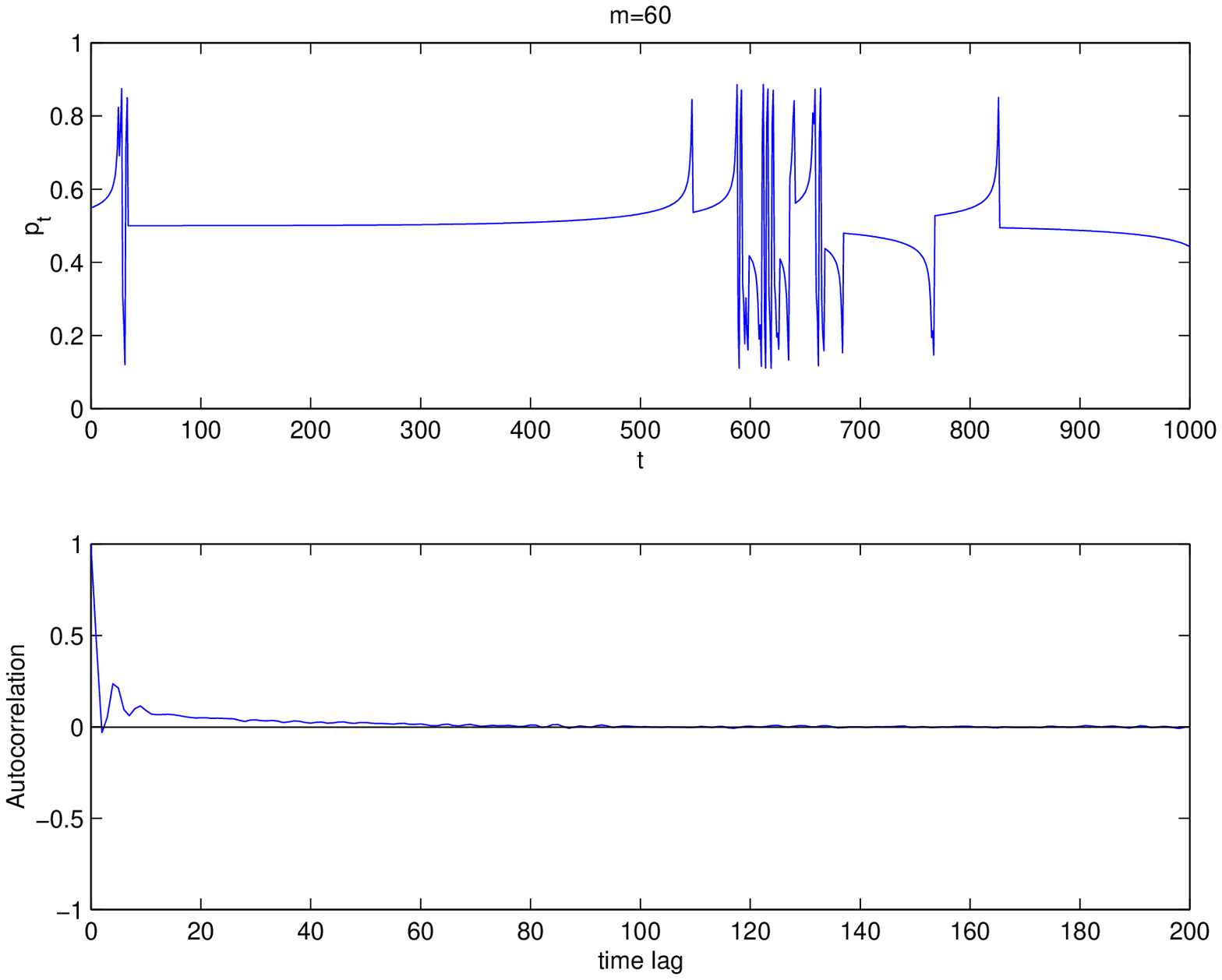}
\includegraphics[width=8cm]{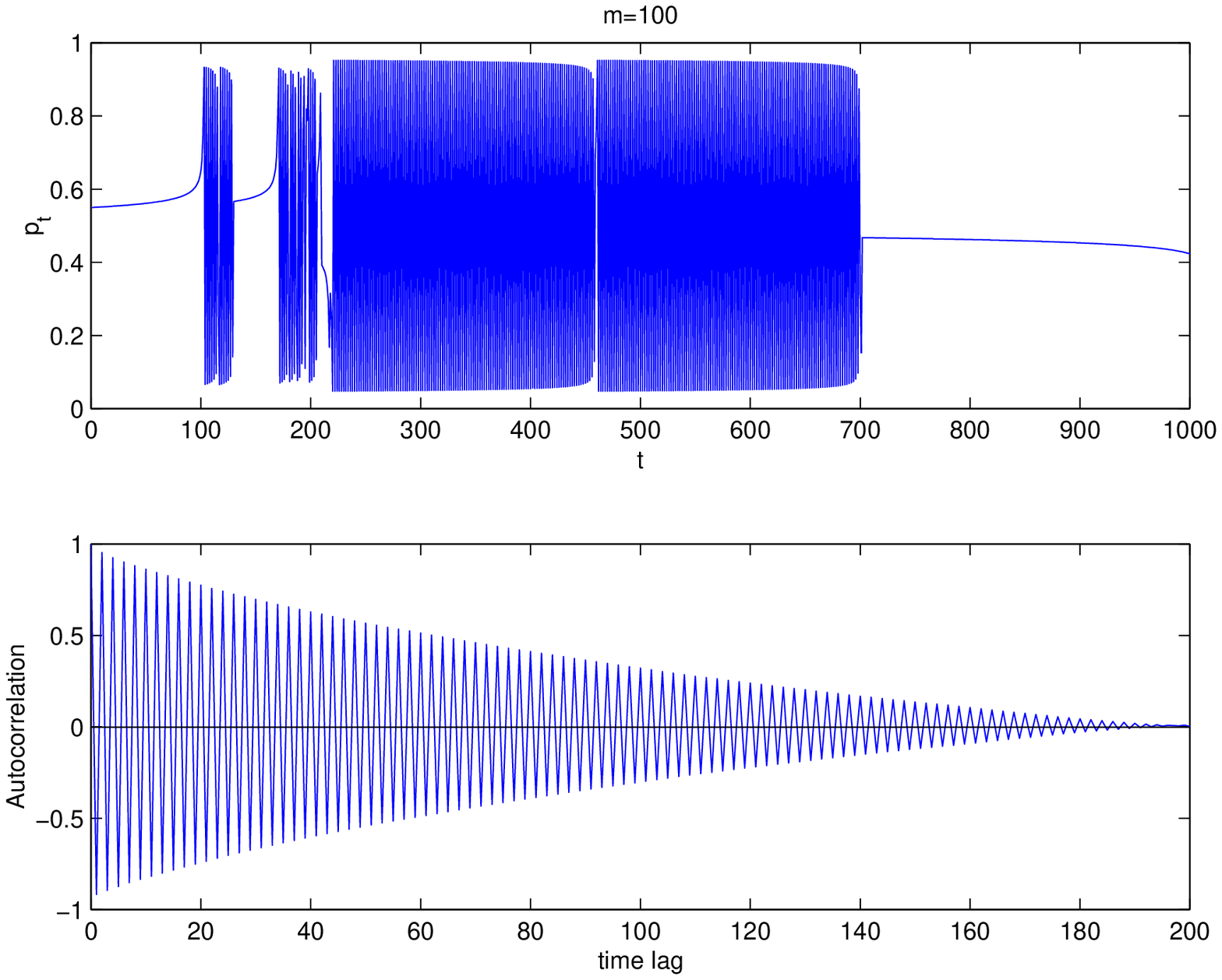}
\caption{\label{fig:7} The upper panels represent the time series $p_t$ for
$m=60$ (left) and $m=100$ (right). The
lower panels represents the corresponding autocorrelation function of
$r_t \propto p-1/2$ for $m=60$ (left) and $m=100$ (right) with the same
parameters $\rho_{hb}=\rho_{bh}=0.72$ and  $\rho_{hh}=\rho_{bb}=0.85$.}
\end{center}
\end{figure}

Figure \ref{fig:7} shows the behavior of the
autocorrelation function for $m=60$ and $m=100$, with the same
values of the other parameters
$\rho_{hb}=\rho_{bh}=0.72$ and $\rho_{hh}=\rho_{bb}=0.85$.
For $m=100$, the presence of the weakly unstable orbits is felt much stronger,
which is reflected in 1) a very strong periodic component of the correlation 
function and 2) its slow decay. Even for $m=60$, the correlation function
does not decay fast enough compared to the typical duration of speculative
bubbles to be in quantitative agreement with empirical data. This anomalously
large correlation of the returns is obviously related to the deterministic
dynamics of the returns. We thus expect that including stochastic noise
due to a finite number $N$ of agents (see below) and adding external noise
due to ``news'' will whiten $r_t$ significantly.

\begin{figure}
\begin{center}
\includegraphics[width=15cm]{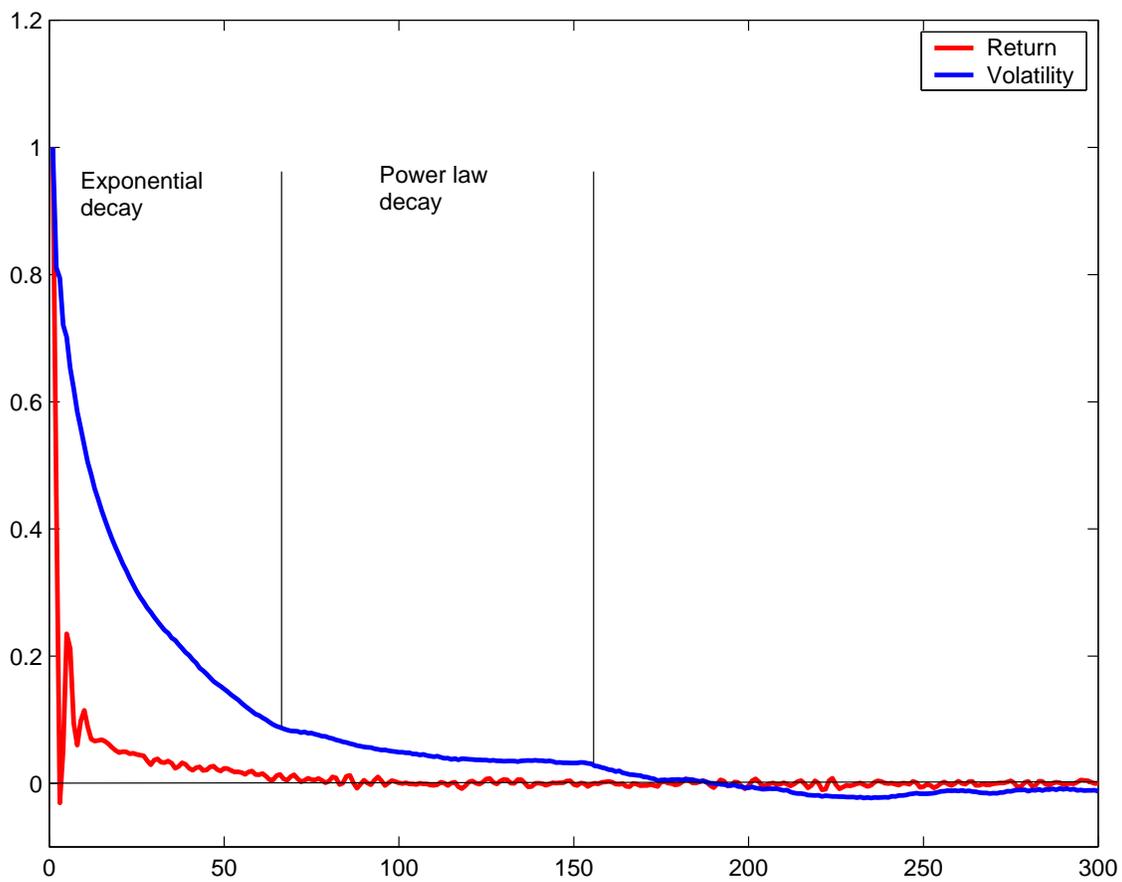}
\caption{\label{fig:8} Autocorrelation function of the returns and of
the volatility for $m=60$ polled agents and the parameters
$\rho_{hb}=\rho_{bh}=0.72$ and  $\rho_{hh}=\rho_{bb}=0.85$.}
\end{center}
\end{figure}

Figure \ref{fig:8} compares the correlation function for
the returns time series $r_t \propto p_t-1/2$ and the volatility time series
defined as $|r_t|$. The volatility is an important measure of risks and
thus plays an important role in portfolio managements and option pricing and
hedging. Note that taking the absolute value of the return removes the
one source of irregularity stemming from the change of sign of $r_t 
\propto p_t-1/2$
to focus on the local amplitudes. We observe in figure \ref{fig:8}
a significantly longer correlation time for the volatility. 
Moreover, the correlation function of the volatility first decays exponentially
and then as a power law. This behavior has previously been documented 
in many econometric works (Ding {\it et al.} (1993), Ding and Granger (1996), 
M\"uller {\it et al.} (1997), Dacorogna {\it et al.} (1998),
Arneodo {\it et al.} (1998), Ballocchi {\it et al.} (1999), Muzy {\it et al.} (2001)).

\section{Asymmetric cases}

We have seen that the symmetric case $\rho_{hb}=\rho_{bh}$ and 
$\rho_{hh}=\rho_{bb}$ is plagued by the weakly unstable periodic
orbits which put a strong and unrealistic imprint on the statistical
properties of the return time series. It is natural to
argue that breaking the symmetry will destroy the strength of these
periodic orbits.

 From a behavioral point of view, it is also quite clear
that the attitude of an investor is not symmetric. One can expect a priori
a stronger bullish buy-and-hold index $\rho_{hb}-1/2$ than bearish
buy-and-hold index $\rho_{bh}-1/2$: one is a priori more prone to hold
a position in a bullish market than in a bearish one. Similarly, we
expect a smaller bullish reversal index $1-\rho_{hh}$ than bearish
reversal index $1-\rho_{bb}$: speculative bubbles are rarely seen on
downward trends as it is much more common that increasing prices are
favorably perceived and can be sustained much longer without reference
to the fundamental price.

Such an asymmetry has been clearly demonstrated empirically
in the difference between
the rate of occurrence and size of extreme drawdowns compared to drawups in stock market
time series (Johansen and Sornette (2001)). Drawdowns (drawups) are
defined as the cumulative losses (gains) from the last local maximum
(minimum) to the next local minimum (maximum). Drawdowns and drawups are
very interesting because they offer
a more natural measure of real market risks than the variance, the 
value-at-risk
or other measures based on fixed time scale distributions of returns.
For the major stock market indices,
there are very large drawdowns which are ``outliers'' while drawups do not
exhibit such drastic change of regime. For major companies, drawups
of amplitude larger than $15\%$ occur at a rate about twice as large as
the rate of drawdowns, but with lower absolute amplitude.

\begin{figure}
\begin{center}
\includegraphics[width=15cm]{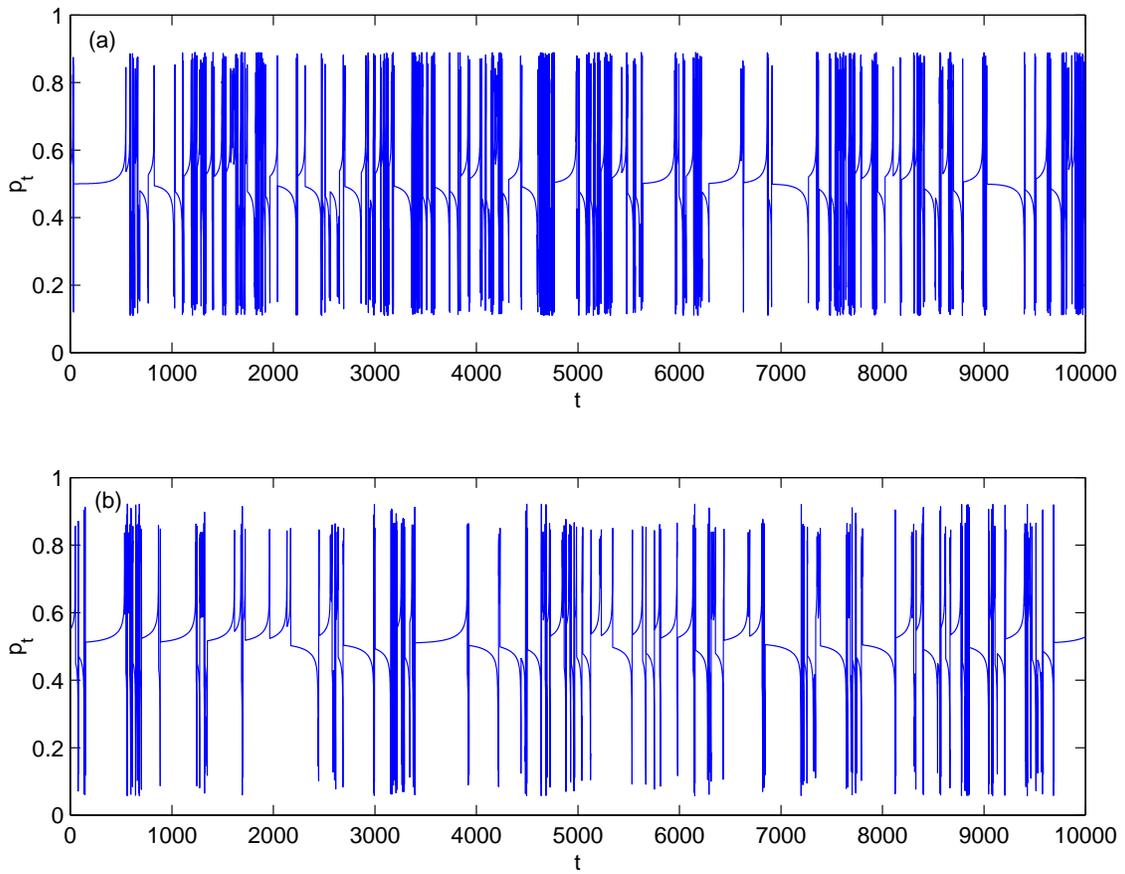}
\caption{\label{fig:8bis} Time evolution of $p_t$ over $10000$ time steps
for $m=60$ polled agents in (a) a symmetric case $\rho_{hb}=\rho_{bh}=0.72$ and
$\rho_{hh}=\rho_{bb}=0.85$ and (b) an asymmetric case 
$\rho_{hb}=0.72$, $\rho_{bh}=0.74$,
$\rho_{hh}= 0.85$ and $\rho_{bb}=0.87$.}
\end{center}
\end{figure}

Figure \ref{fig:8bis} compares the dynamics for the symmetric system (upper panel (a))
and for the asymmetric system (lower panel (b)). It is clear that, as expected,
the number of periodic orbits decreases significantly in the asymmetric
system. However, there are still an unrealistic number of negative bubbles.
It is not possible to increase the asymmetry sufficiently strongly without
exiting from the chaotic regime. This unrealistic feature is thus an intrinsic
property and limitation of the present model. We shall indicate in 
the conclusion
possible extensions and remedies.

\begin{figure}
\begin{center}
\includegraphics[width=15cm]{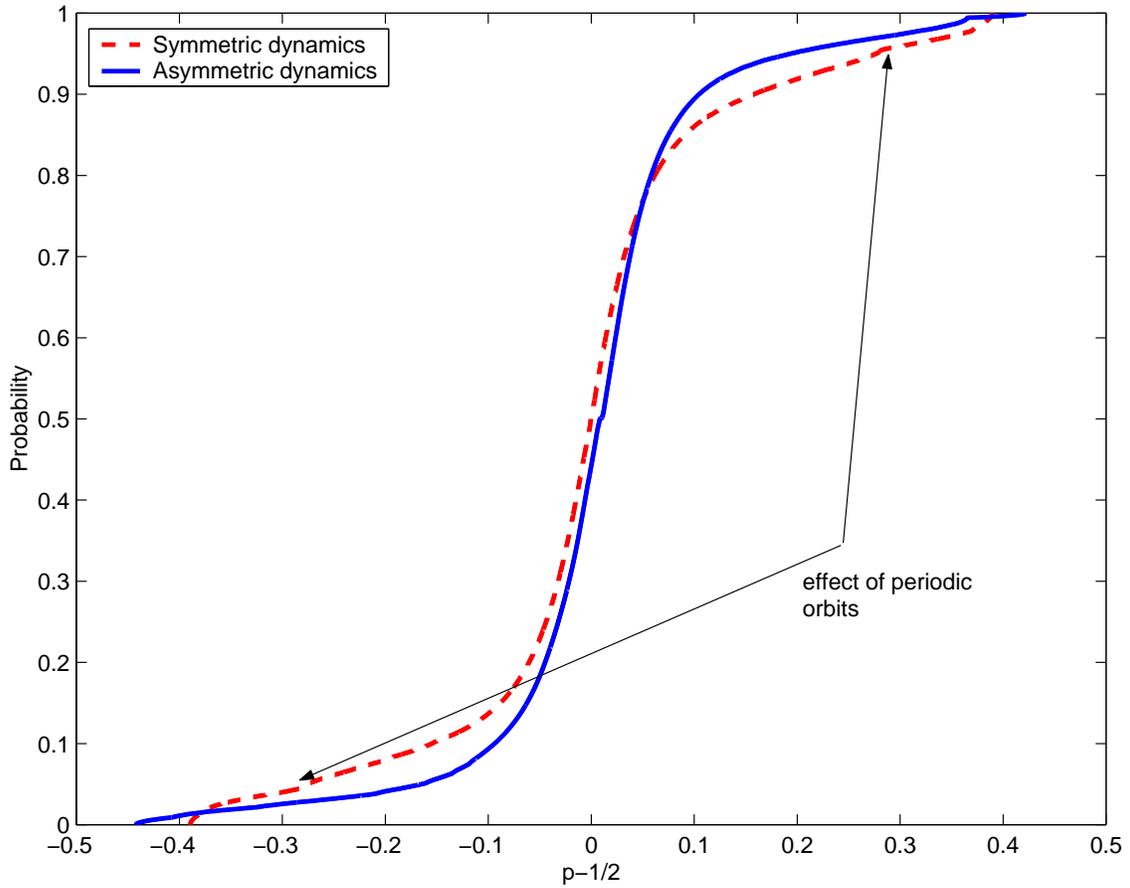}
\caption{\label{fig:9} Distribution function of $p-1/2$ for $m=60$ polled
agents and parameters $\rho_{hb}=\rho_{bh}=0.72$ and
$\rho_{hh}=\rho_{bb}=0.85$ (dashed line) and $\rho_{hb}=0.72$, 
$\rho_{bh}=0.74$,
$\rho_{hh}= 0.85$ and $\rho_{bb}=0.87$ (continuous line).}
\end{center}
\end{figure}

Figure \ref{fig:9} compares the cumulative distributions
of $p-1/2$ for $m=60$ for the symmetric and asymmetric cases. The strong
effect of the weakly unstable periodic orbits observed in the periodic case
has disappeared. In addition, the tail of the distribution decays faster in the
asymmetric case, in better (but still not very good) agreement with 
empirical data.

\begin{figure}
\begin{center}
\includegraphics[width=15cm]{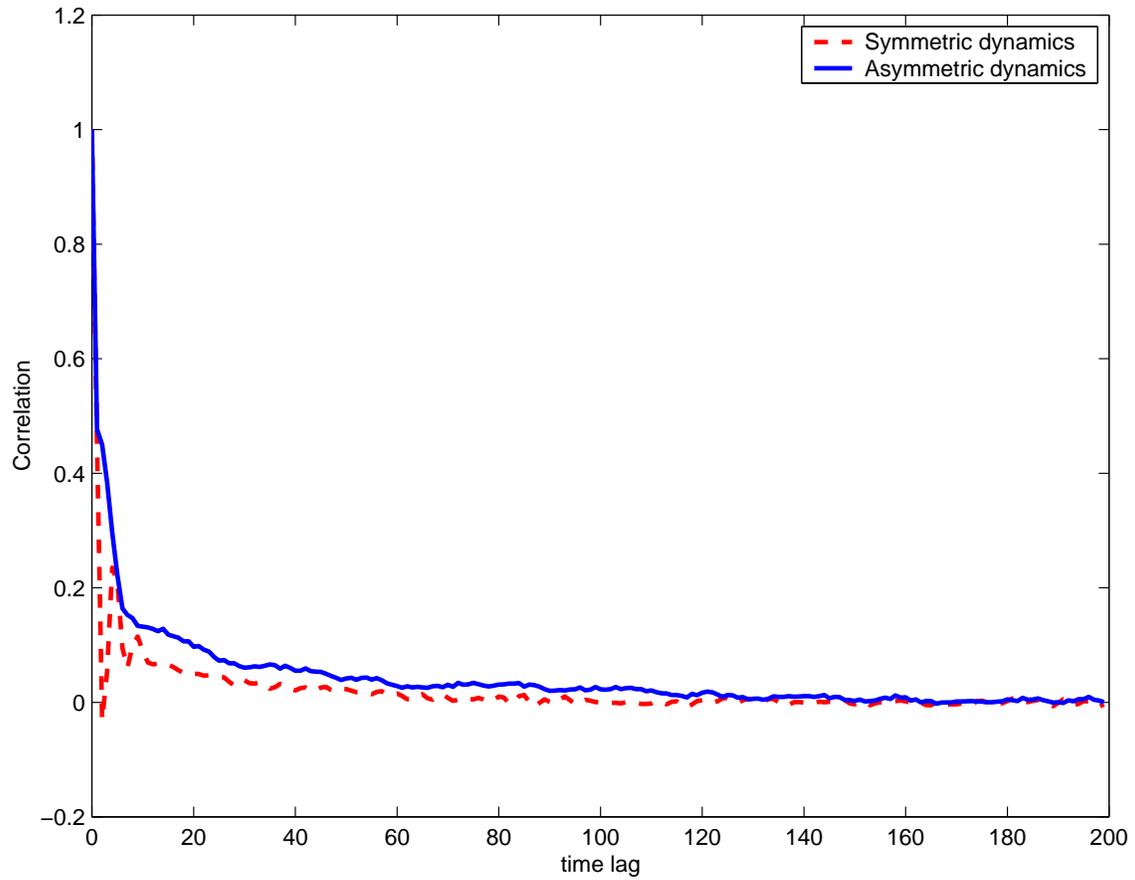}
\caption{\label{fig:10} Correlation function for $m=60$
polled agents and parameters $\rho_{hb}=\rho_{bh}=0.72$ and
$\rho_{hh}=\rho_{bb}=0.85$ (dashed line) and $\rho_{hb}=0.72$, 
$\rho_{bh}=0.74$,
$\rho_{hh}= 0.85$ and $\rho_{bb}=0.87$ (continuous line).}
\end{center}
\end{figure}

Figure \ref{fig:10} shows the correlation function of the returns for a
symmetric and an asymmetric case. In the asymmetric case, there is no trace
of oscillations but the decay is slightly slower.

\section{Finite size effects}

Until now, our analysis has focused on the limit of an infinite 
number $N \to \infty$
of agents, in which each agent polls randomly $m$ agents among $N$. In this
limit, we have shown that, for a large domain in the parameter space, 
the dynamics
of the returns is chaotic with interesting and qualitatively 
realistic properties.

\subsection{Finite-size effects in other models}

We now investigate finite-size effects resulting from a finite number $N$ of
interacting agents trading on the stock market. This issue of the role of
the number of agents has recently been investigated
vigorously with surprising results.
First, Egenter {\it et al.}, (1999) studied the $N$-dependence of the dynamical
properties of price time series of the Kim-Markowitz (1989) and of the Lux-Marchesi (1999) 
models. They found that, if this number $N$ goes to infinity, nearly periodic
oscillations occur and the statistical properties of the price time series
become completely unrealistic. Stauffer (1999) reviewed this work
and others such as the Levy-Levy-Solomon (1995, 2000) model:
realistically looking price
fluctuations are obtained for $N \sim 10^2$, but for $N \sim 10^6$ the prices
vary smoothly in a nearly periodic and thus unrealistic way.
The model proposed by Farmer (1998) suffers from the same problem:
with a few hundred investors,
most investors are fundamentalists during calm times, but bursts of
high volatility coincide with large fractions of noise traders. When 
$N$ becomes
much larger, the fraction of noise traders goes to zero
in contradiction to reality. On a somewhat different issue, Huang and 
Solomon (2001)
have studied finite-size effects in dynamical systems of price evolution with
multiplicative noise. They find that the
exponent of the Pareto law obtained in stochastic multiplicative 
market models is crucially
affected by a finite $N$ and may cause in the absence of an appropriate social
policy extreme wealth inequality and market instability.
Another model (apart from ours) where the market may stay realistic even for $N
\rightarrow \infty$ seems to be the Cont-Bouchaud percolation model 
(2001). However,
this only occurs for an unrealistic tuning of the percolation 
concentration to its
critical value.
Thus, in most cases, the limit
$N \rightarrow \infty$ leads to a behavior of the simulated markets which becomes quite
smooth or periodic and thus predictable, in contrast to real markets. Our model
which remains (deterministically) chaotic is thus a significant improvement
upon this behavior. We trace this improvement on the highly nonlinear behavior
resulting from the interplay between the imitative and contrarian behavior. 
It has thus been argued (Stauffer (1999)) that,
if these previous models are good
descriptions of markets, then real markets with their strong random
fluctuations are dominated by a rather limited number of large players:
this amounts to assume that the hundred most important
investors or investment companies have much more influence than the millions
of less wealthy private investors.

There is another class of models, the minority games (Challet and Zhang (1997)), in which the 
dynamics remains complex even in the limit $N \to \infty$. It has been
established that the fluctuations of the sum of the aggregate demand 
have an approximate scaling with similar sized fluctuations (volatility/standard
deviation) for any $N$ and $m$ for the scale scaled variable
$2^m/N$, where $m$ is the memory length (Challet {\it et al.} (2000)). In 
a generalization, the so-called Grand Canonical version of the Minority Game
(Jefferies {\it et al.} (2001)), where the agents
have a confidence threshold that prevents them from playing if their
strategies have not been successful over the last $T$ turns,
the dynamics can depend more
sensitively on $N$: as $N$ becomes small, the dynamics can become 
quite different.
For large $N$, the complexity remains.

The difference between the limit $N \to \infty$ considered up to now in this paper
and the case of finite $N$ is that $p_t$ is no more the fraction of bullish agents. 
For finite $N$, $p_t$ must be interpreted as 
the probability for an agent to be bullish. Of course, in the limit of large $N$,
the law of large numbers ensures that the fraction of bullish agents becomes
equal to the probability for an agent to be bullish. There are several ways to
implement a finite-size effect. We here discuss only the two simplest ones.

\subsection{Finite external sampling of an infinite system}

Consider a system with an infinite number of agents for which the 
fraction $p_t$
of bullish agents is governed by the
deterministic dynamics (\ref{anne}). At each time step $t$,
let us sample a finite number $N$ of them
to determine the fraction of bullish agents. We get a number $n$,
which is in general close
but not exactly equal to $N p_t$ due to statistical fluctuations. 
The probability to find
$n$ bullish agents among $N$ agents is indeed given by the binomial law
\be
\Pr(n) = {N \choose n} p^n (1-p)^{N-n}~.
\label{pbino}
\ee
This shows that the observed proportion $\tilde p = n/N$ of bullish
agents is asymptotically normal with mean $p$ and standard deviation
$1/\sqrt{p(1-p) N}$~: $\Pr( \tilde p) \sim {\cal N}(p,1/\sqrt{p(1-p) N})$. 
Iterating the sampling among $N$ agents at each time step gives
a noisy dynamics $\tilde p_t$ shadowing the true deterministic one.

\begin{figure}
\begin{center}
\includegraphics[width=15cm]{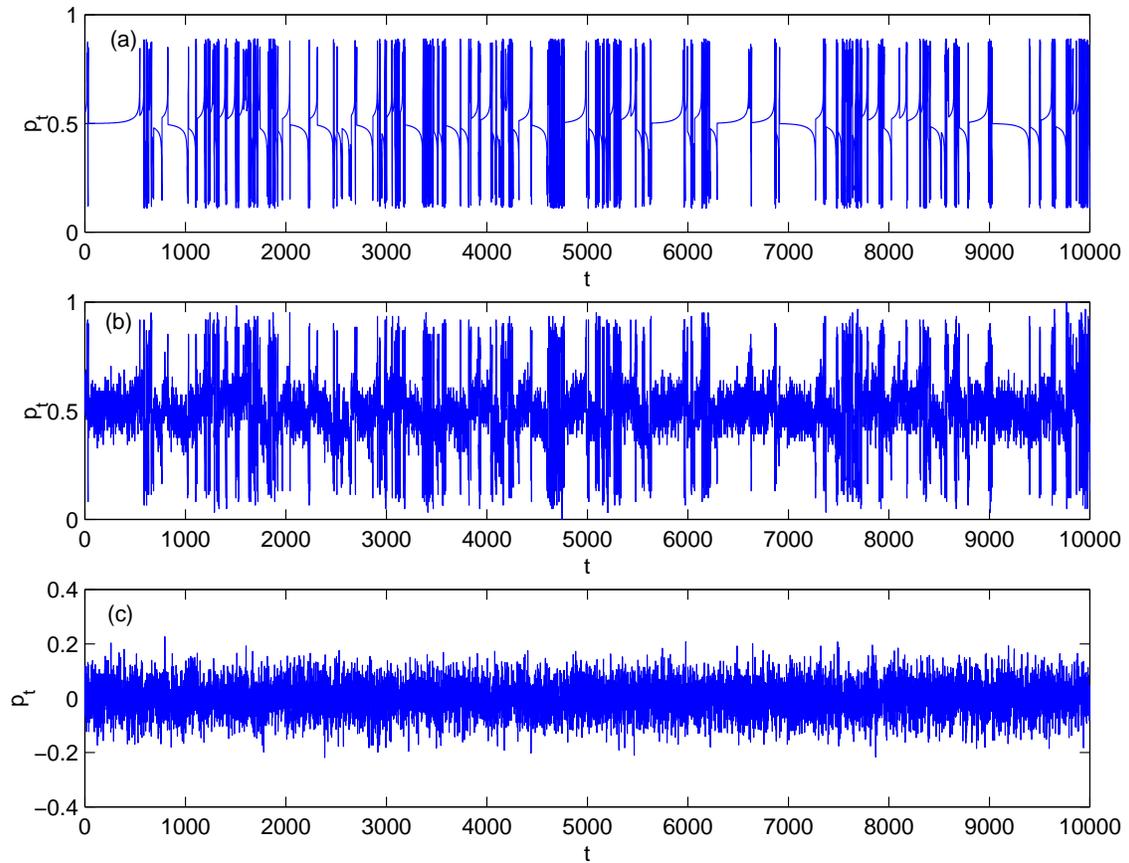}
\caption{\label{fig:14} Time evolution of $p_t$  over $10000$ time steps
for $m=60$ polled agents with (a) $N=\infty$, (b) $N=m+1=61$ agents 
and parameters
$\rho_{hb}=\rho_{bh}=0.72$ and  $\rho_{hh}=\rho_{bb}=0.85$. The panel
(c) represents the noise due to the finite size of the system and is obtained
by subtracting the time series in panel (a) from the time series in 
panel (b).}
\end{center}
\end{figure}

Figure \ref{fig:14} compares the dynamics of the deterministic $p_t$ 
corresponding to $N \to \infty$ (panel (a)) with $\tilde p_t$ for 
a number $N=m+1=61$ of sampled agents among the infinite ensemble of them (panel (b)). 
Panel (c) is the ``noise'' time series defined as  $\tilde p_{t} - p_t$, {\it i.e}, by
subtracting the time series of panel (a) from the time series of panel (b). 
The noise time series of panel (c) thus represents the statistical fluctuations
due to the finite sampling of agents'opinions. Figure \ref{fig:14}-b shows the
characteristic
volatility clusters which is one of the most important stylized properties
of empirical time series.

\begin{figure}
\begin{center}
\includegraphics[width=15cm]{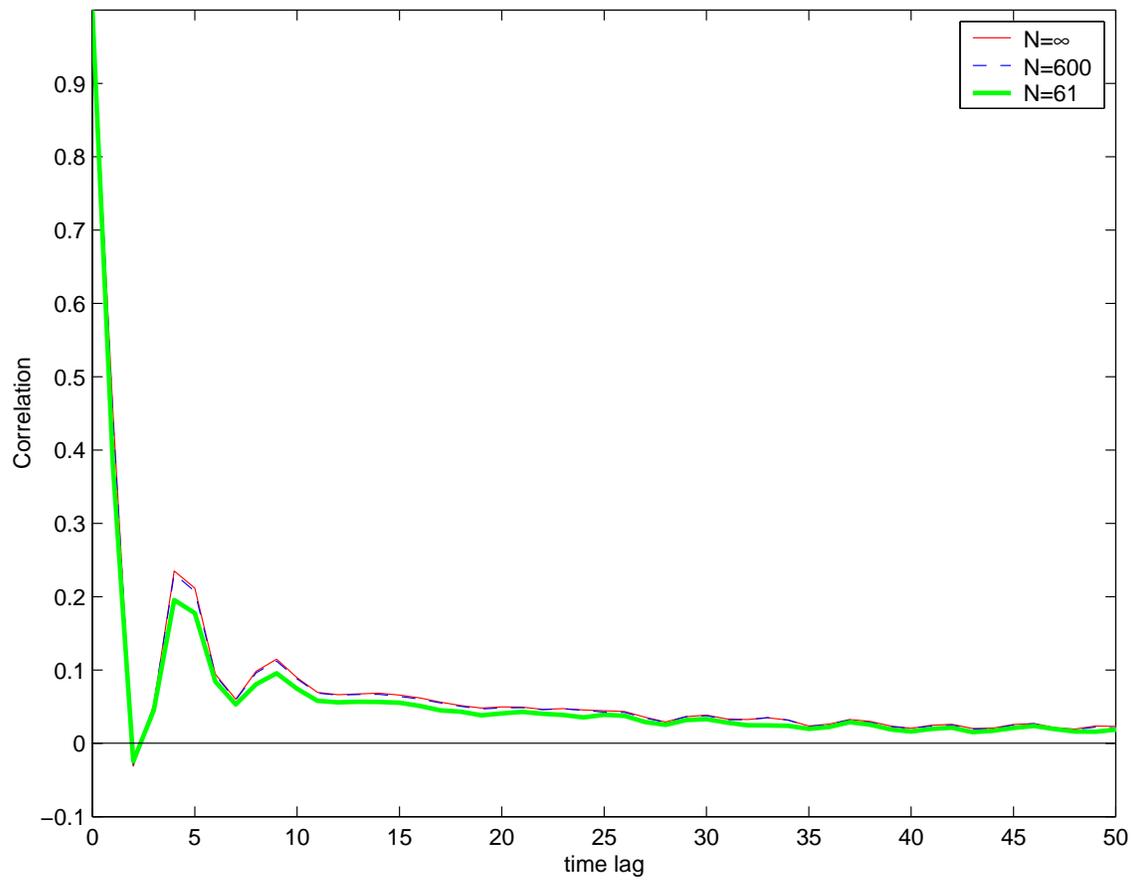}
\caption{\label{fig:15} Correlation function for $m=60$ polled
agents with  $N=\infty$ (thin line), $N=600$ (dashed line) and $N=61$
(continuous line) agents  and parameters
$\rho_{hb}=\rho_{bh}=0.72$ and  $\rho_{hh}=\rho_{bb}=0.85$.}
\end{center}
\end{figure}

For large $N$, we can write
\be
{\tilde p}_t = p_t + \frac{1}{\sqrt{p_t(1-p_t) N}}W_t
\ee
where $\{W_t\}$ are iid gaussian variables with zero mean and unit
variance.
Therefore, the correlation function Corr$_N(\tau)$ at lag
$\tau \neq 0$ is obtained from that for $N \to \infty$ by multiplication
by a constant factor:
\bea
\mbox{Corr}_N(\tau)&=&\frac{N  \mbox{Var}(p)}{{\rm E}[1/\{p(1-p)\}]+N 
\mbox{Var}(p)}
\times \mbox{Corr}_\infty(\tau)~~~~~ \mbox{and}~\tau \neq 0,\\
&\simeq& \mbox{Corr}_\infty(\tau)~~~~~ \mbox{for large}~N~,
\eea
where ${\rm E}[x]$ denotes the expectation of $x$ with respect to the continuous
invariant measure of the dynamical system (\ref{anne}). Note that 
${\rm E}[1/\{p(1-p)\}]$ always exists for $m<\infty$ since the support of the
continuous measure of (\ref{anne}) with respect to Lebesgue measure is bounded
from below by a value strictly larger than $0$ and from above by a value 
strictly less than $1$.
Figure \ref{fig:15} shows that the correlation function of $\tilde p_t$
is very close to that of the deterministic trajectory $p_t$.

\begin{figure}
\begin{center}
\includegraphics[width=15cm]{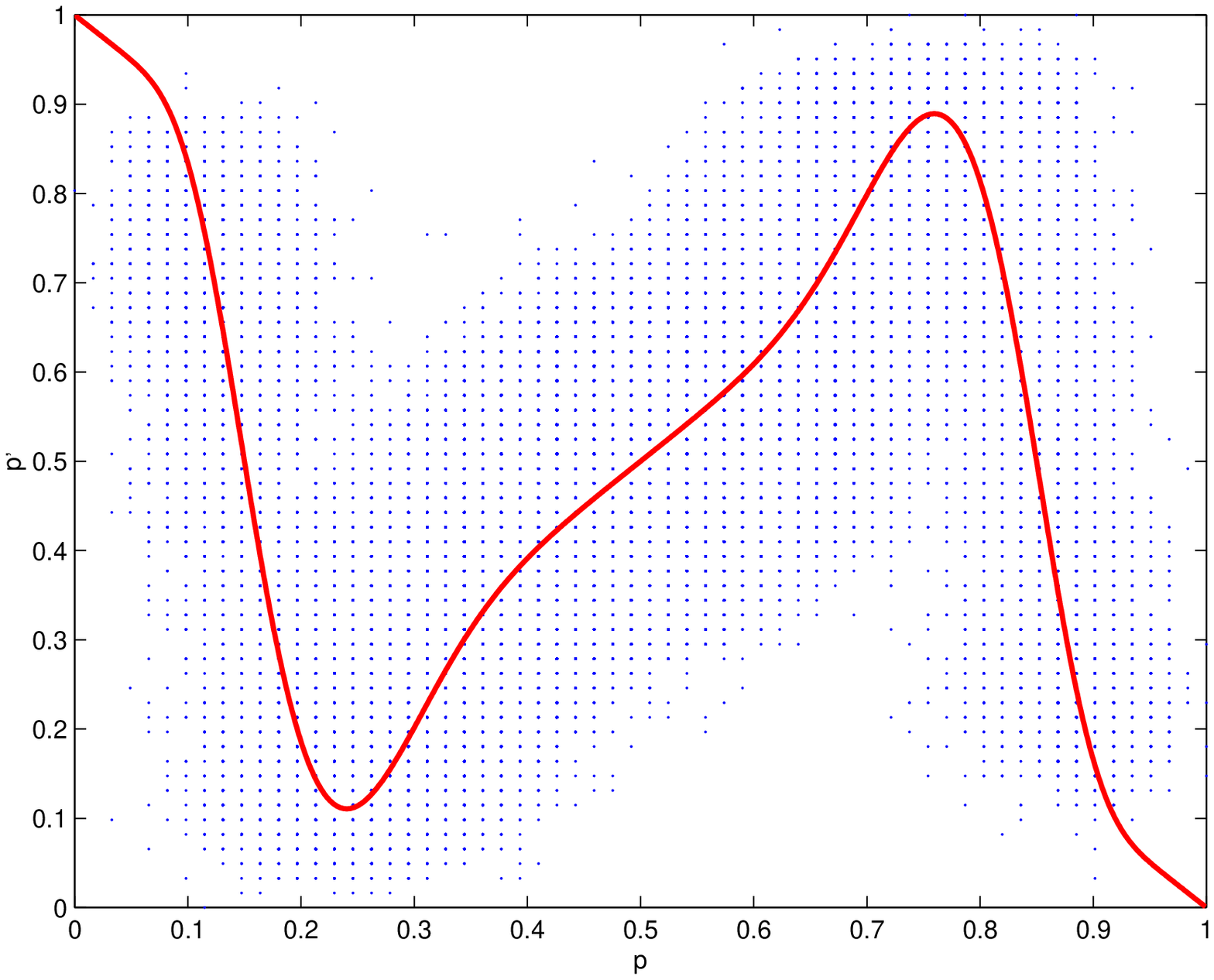}
\caption{\label{fig:16} Return map of the fraction of bullish
agents for $m=60$ polled agents among
$N=61$ agents (points) and the deterministic trajectory (continuous
line) corresponding to $N=\infty$ agents. 
The parameters are $\rho_{hb}=\rho_{bh}=0.72$ and
$\rho_{hh}=\rho_{bb}=0.85$.}
\end{center}
\end{figure}

\begin{figure}
\begin{center}
\includegraphics[width=15cm]{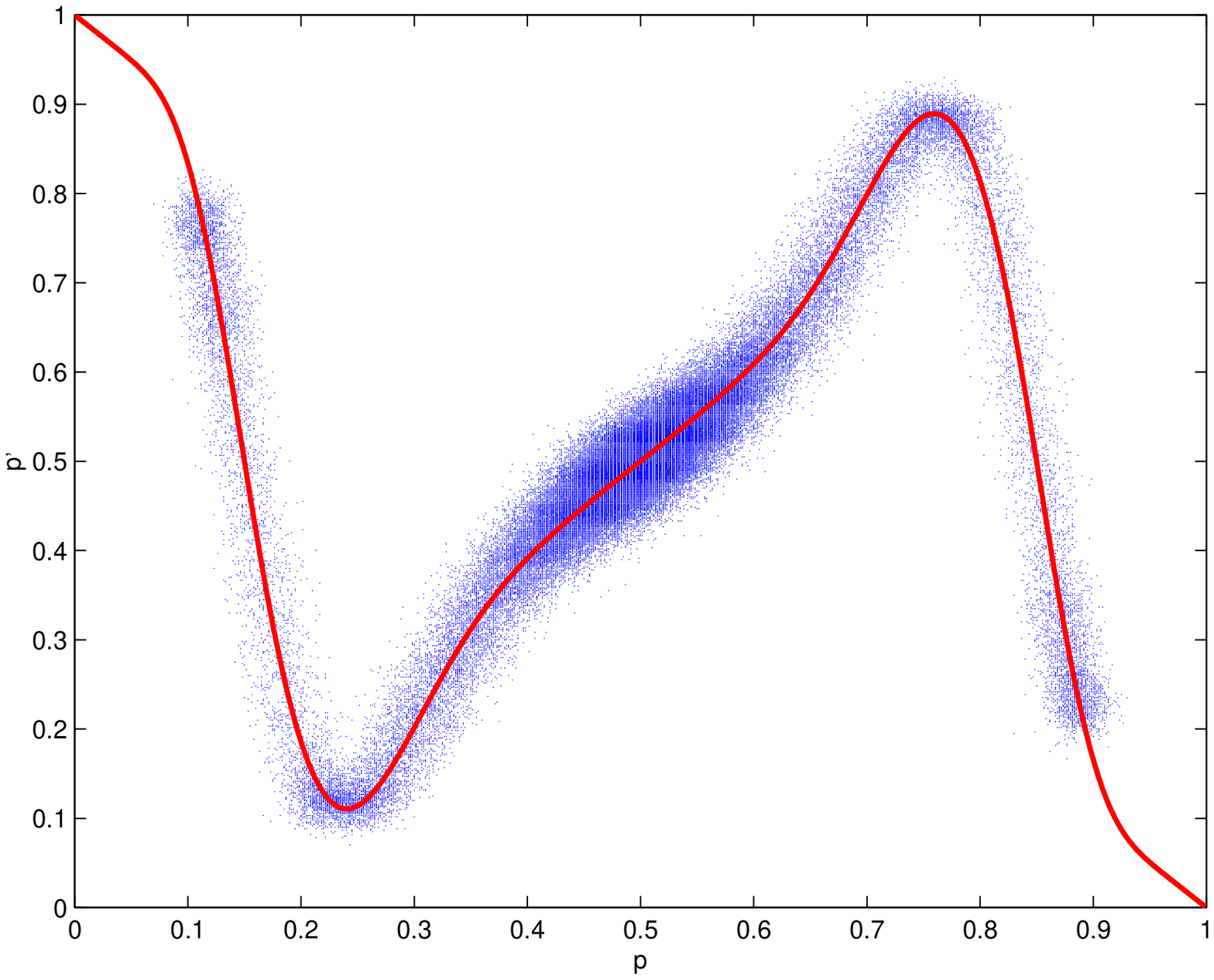}
\caption{\label{fig:17} Return map of the fraction of bullish
agents for $m=60$ polled agents among
$N=600$ agents (points) and the deterministic trajectory (continuous
line) corresponding to $N=\infty$ agents. 
The parameters are $\rho_{hb}=\rho_{bh}=0.72$ and $\rho_{hh}=\rho_{bb}=0.85$.}
\end{center}
\end{figure}

To quantify further the impact of the statistical noise
stemming from the finite size of the market,
figures \ref{fig:16} and \ref{fig:17} show the return maps of $\tilde p_t$,
{\it i.e}, $\tilde p_{t+1}$ as a function of $\tilde p_t$,
for $m=60$ polled agents among a total number $N=61$
of agents (fig \ref{fig:16}) and $N=600$ (fig \ref{fig:17}).

\begin{figure}
\begin{center}
\includegraphics[width=15cm]{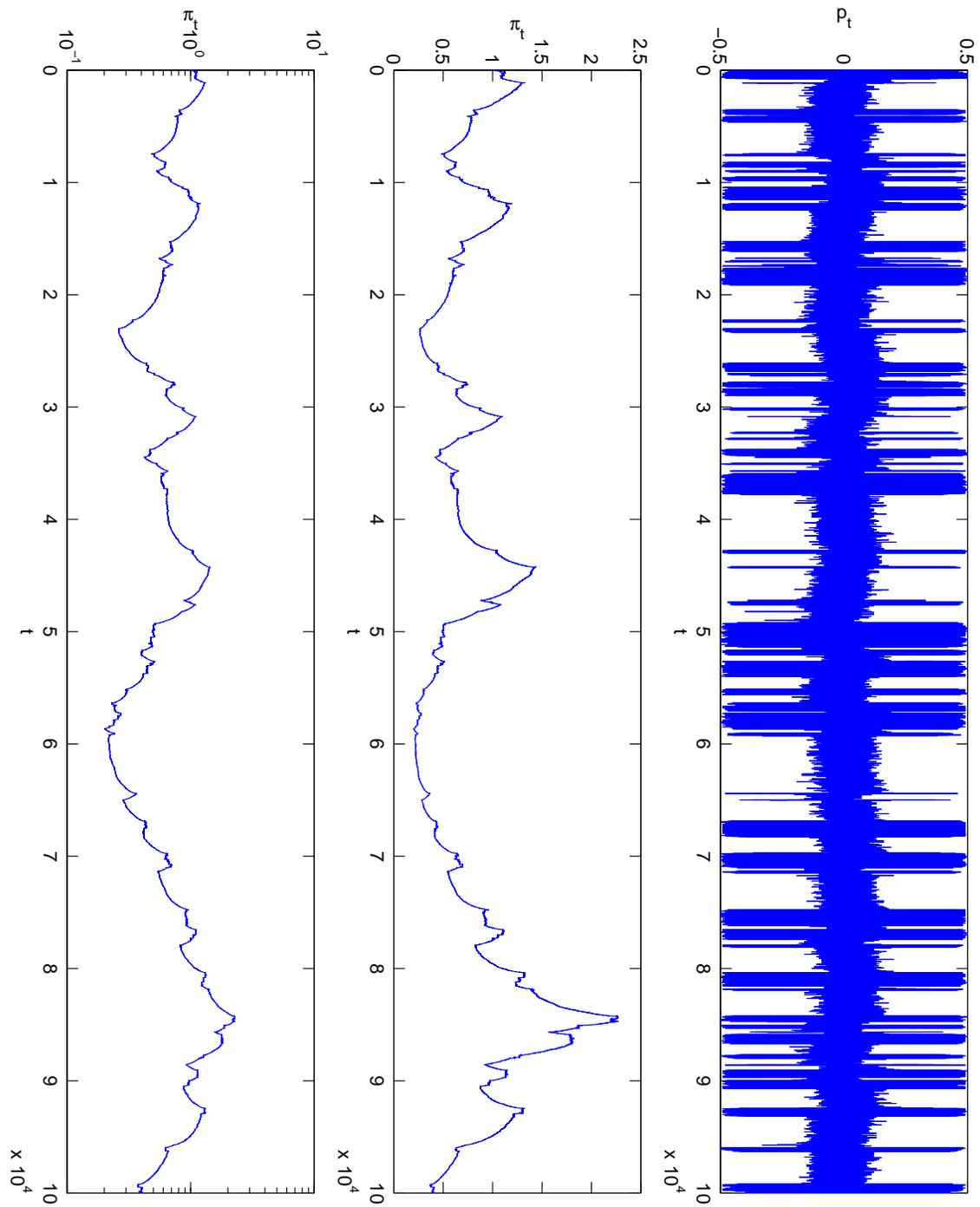}
\caption{\label{fig:19} Upper panel: return trajectory ${\tilde r}_t = \gamma
{\tilde p}_t -1/2$ for $m=100$, $N=100$, $\rho_{hb}=\rho_{bh}=0.72$ and
$\rho_{hh}=\rho_{bb}=0.85$ and $\gamma =0.01$. Middle panel: price trajectory
obtained by
$\pi_t = \pi_{t-1}~\exp [ {\tilde r}_t]$~. Lower panel: same as the 
middle panel
with $\pi_t$ shown in logarithmic scale. Note the ``flat trough-sharp peak''
structure of the log-price trajectory (Roehner Sornette (1998)).}
\end{center}
\end{figure}

Figure \ref{fig:19} shows the price trajectory
obtained by $\pi_t = \pi_{t-1}~\exp [ {\tilde r}_t]$ in linear
and logarithmic scale. The super-exponential acceleration of the price
giving rise to sharp peaks in the semi-logarithmic
representation (Roehner and Sornette (1998)) is clearly 
visible.

\subsection{Finite number $N$ of agents}

We now introduce a genuine finite stock market with $N$ agents.
We assume that the agents do not know the exact number $N$ of agents
in the market (this is realistic) and they are
in contact with only $m$ other agents that they poll
at each time period. Not knowing the true value of $N$ but
assuming it to be large, it is rational for them to develop the best
predictor of the dynamics by assuming the ideal case of an infinite
number of agents with $m$ polled agents and thus use the deterministic dynamics
(\ref{anne}) as their best predictor.

At each time period $t$, each agent thus chooses randomly $m$ agents
that she polls. She then counts the number of bullish and bearish agents among her polled
sample of $m$ agents. This number divided by $ 
m$ gives her an estimation
${\hat p}_t$ of the probability $p_t$ be to bullish at time $t$.
Introducing this estimation in the deterministic equation 
(\ref{anne}), the agent
obtains a forecast $\hat p'$ of the true probability $p'$ to be bullish at the
next time step.

\begin{figure}
\begin{center}
\includegraphics[width=15cm]{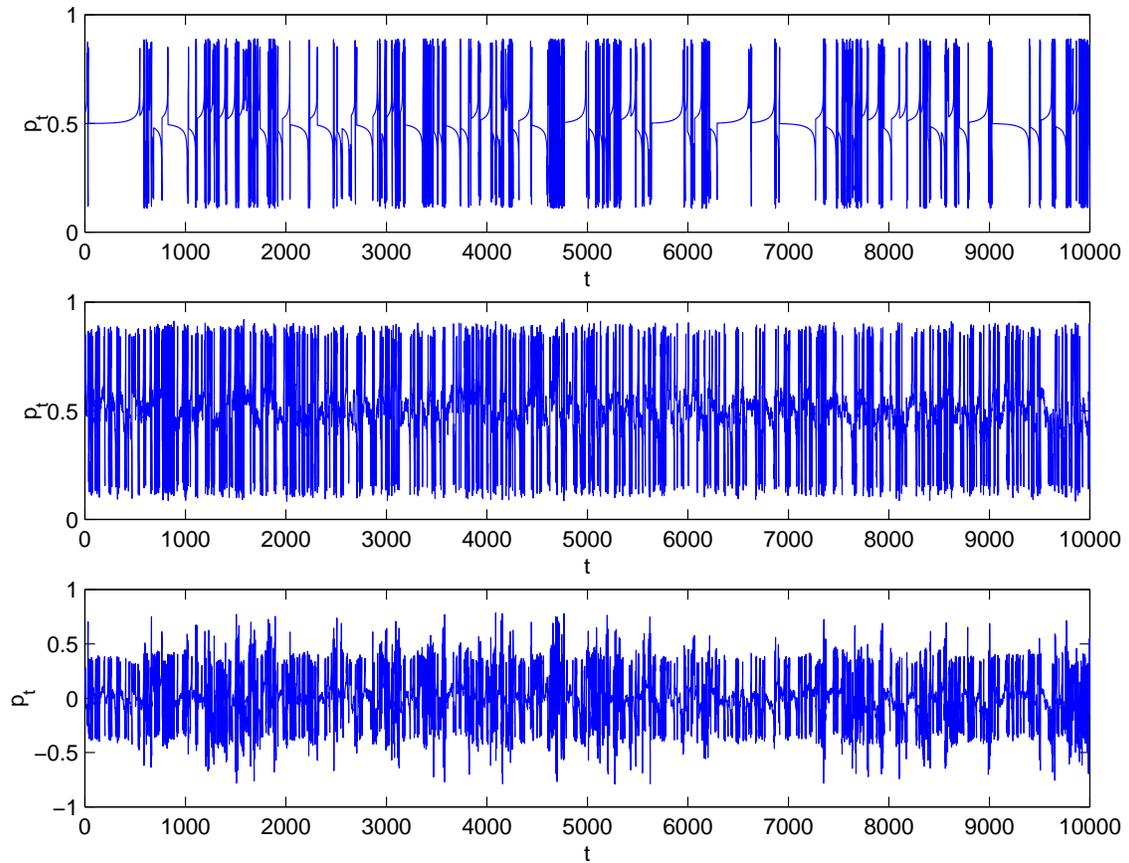}
\caption{\label{fig:20} Evolution of the system over $10000$ time steps
for $m=60$ polled agents with (upper panel) $N=\infty$, (second 
panel) $N=m+1=61$ and parameters
$\rho_{hb}=\rho_{bh}=0.72$ and $\rho_{hh}=\rho_{bb}=0.85$. The
lower panel represents the ``noise'' introduced by the finite size of 
the system and
is obtained by subtracting the upper panel from the second panel.}
\end{center}
\end{figure}

Results of the simulations of this model are shown in figure \ref{fig:20}.
We observe a significantly stronger ``noise'' compared to the previous
section, which is expected since the noise is itself injected in the dynamical
equation at each time step. As a consequence, the correlation function of the
returns and of the volatility decay faster than their deterministic 
counterpart.
The correlation of the volatility still decays about ten times slower than
the correlation of the returns, but this clustering of volatility is not
sufficiently strong compared to empirical facts.

Other more realistic models of a finite number of agents can be 
introduced. For instance,
at time $t$, consider an agent among the $N$.
She chooses $m$ other agents randomly and polls them. Each of them is either
bullish or bearish as a result of decisions taken during the previous 
time period.
She then counts the number of bullish agents among the $m$, and then determines
her new attitude using the rules (\ref{rules}). If she is polled at time $t+1$
by another agent, her attitude will be the one determined from $t$ to $t+1$.
In this way, we never refer to the deterministic dynamics $p_t$ but only
to its underlying rules. As a consequence,
this deterministic dynamics does not exert an attraction that 
minimizes the effect
of statistical fluctuations due to finite sizes. This approach is similar
to going from a Fokker-Planck equation (equation (\ref{anne})) to a Langevin
equation with finite-size effects. This class of
models will be investigated elsewhere.

\section{Conclusions}

The traditional concept of stock market dynamics envisions a stream of
stochastic ``news'' that may move prices in random directions. This paper, in
contrast, demonstrates that certain types of deterministic behavior---mimicry
and contradictory behavior alone---can already lead to chaotic prices.

If the market prices are assumed to follow the $p_{t}$ behavior, our
description refers to the well-known evolution of the speculative bubbles. Such
apparent regularities often occur in the stock market and form the basis of the
so-called ``technical analysis'' whereby traders attempt to predict future
price
movements by extrapolating certain patterns from recent historical prices. Our
model provides an explanation of birth, life and death of the
speculative bubbles in this context.

While the traditional theory of rational anticipations exhibits and
emphasizes
self-re\-in\-forcing mechanisms, without either predicting their inception
nor their collapse, the strength of our model is to justify the occurrence of
speculative bubbles. It allows for their collapse by taking into account the
combination of mimetic and antagonistic behavior in the formation of 
expectations
about prices.

The specific feature of the model is to combine these two Keynesian aspects of
speculation and enterprise and to derive from them behavioral rules based on
collective opinion: the agents can adopt an imitative and gregarious
behavior,
or, on the contrary, anticipate a reversal of tendency, thereby detaching
themselves from the current trend. It is this duality, the continuous
coexistence of these two elements, which is at the origin of the properties
of our model: chaotic behavior and the generation of bubbles.

It is a common wisdom that deterministic chaos leads to 
fundamental limits of predictability because the tiny inevitable
fluctuations in 
those chaotic systems quickly snowball in unpredictable ways.
This has been investigated in
relation with for instance long-term weather patterns. 
However, in the context of our models,
we have shown
that the chaotic dynamics of the returns alone cannot be 
the limiting factor
for predictability, as it contains too much residual correlations.
Endogenous fluctuations due to finite-size effects
and external news (noise) seem to be needed as important factors leading to
the observed randomness of stock market prices. The relation between
these extrinsic factors and the intrinsic ones studied in this paper
will be explored
elsewhere.

\vskip 0.5cm
{\bf Remark and Acknowledgements}: This paper is an outgrowth and
extension of unpublished work by three of
us (AC, JPE, AM) which was in turn based on the Ph.D. of Anne Corcos.
We are grateful to J.V. Andersen for useful
discussions. This work was partially supported by the Fonds National Suisse
(JPE and AM) and by the James S. Mc Donnell Foundation 21st century
scientist award/studying complex system (DS).

\pagebreak

\section*{Appendix}

We expand $F_m(p)$ around the fixed point $p=1/2$, so that, using the symmetry
of $F_m(p)$
\be
F_m(p)=\frac{1}{2}+ F_m'(1/2) \cdot \left( p- \frac{1}{2} \right) +
F_m^{(3)}(1/2) \cdot \left( p- \frac{1}{2} \right)^3 + \cdots
\ee

First of all, it is obvious to show by recursion that
\bea
F_m'(1/2) &=& 1 - 2 g_m(1/2)  - g_m'(1/2) \\
F_m^{(2k+1)}(1/2)  &=& -2 (2k+1) g_m^{(2k)}(1/2) - g_m^{(2k+1)}(1/2)~ ~
\mbox{if}~ ~  k>0~.
\eea
The problem thus amounts to calculating the derivatives of $g_m$.

Some simple algebraic manipulations allow to obtain
\bea
g_m'(p)&=&m \sum_{j=0}^{m-1} { {m-1} \choose j} p^{m1-j} (1-p)^j \left[
f \left( \frac{j}{m} \right) - f \left(\frac{j+1}{m} \right) \right] \\
&=& - m \sum_{j=0}^{m-1} { {m-1} \choose j} p^{m1-j} (1-p)^j  \Delta_1 f_m
(j),
\eea
where $\Delta_1 f_m (\cdot)$ is the first order discrete derivative 
of $f \left(
\frac{\cdot}{m} \right)$, which yields
\be
g_m' \left( \frac{1}{2} \right) =- \frac{m}{2^{m-1}}  \sum_{j=0}^{m-1} { {m-1}
\choose j} \Delta_1 f_m (j)~.
\ee

By recursion, it is easy to prove that
\be
g_m^{(k)} \left( \frac{1}{2} \right) = \frac{ (-1)^k~ m !}{2^{m-k} ~k!}
\sum_{j=0}^{m-k} { {m-k} \choose j} \Delta_k f_m (j)
\ee
and $\Delta_k f_m (\cdot)$ is the $k^{th}$ order discrete derivative of $f
\left( \frac{\cdot}{m} \right)$:
\be
\Delta_k f_m (j) = \sum_{i=0}^k {k \choose i} (-1)^i f \left( \frac{j+i}{m}
\right).
\ee

Finally,
\bea
F_m^{(2k+1)}(1/2)&=&\frac{m!}{2^{m-2k-1}~ (2k)!} \left[ \frac{1}{2k+1}
\sum_{j=0}^{m-2k-1} { {m-2k-1} \choose j} \Delta_{2k+1} f_m (j) \right.
\nonumber \\
& & \left.-(2k+1) \sum_{j=0}^{m-2k} { {m-2k} \choose j} \Delta_{2k} f_m (j)
\right].
\eea

\newpage

{\bf References}

\vskip 0.5cm
\parskip5pt\parindent0pt
Arneodo, A., Muzy, J.F. and Sornette, D. (1998),
``Direct'' causal cascade in the stock market, {\it European Physical 
Journal B},
2, 277--282.

Arthur W.B., (1987), `Self-Reinforcing Mechanisms in Economics', {\it 
Center for
Economic Policy Research}, 111, 1--20.

Azariadis C. (1981), `Self Fulfilling Prophecies', {\it Journal of Economic
Theory}, 25(3).

Azariadis C. \& R. Guesnerie (1982), `Proph\'eties autor\'ealisatrices et
persistance des th\'eories', {\it Revue \'economique}, Sept.

Ballocchi, G., M. M. Dacorogna, R. Gencay (1999),
Intraday Statistical Properties of Eurofutures by Barbara Piccinato,
{\it Derivatives Quarterly}, 6, 28-44.

Benhabib J. \& R.H. Day (1981), `Rational Choice and Erratic Behaviour', {\it
Review of Economic Studies}, July, 153.

Bikhchandani S., D. Hirshleifer, I. Welch (1992), `A Theory of Fads, Fashion,
Custom and Cultural Changes as Informational Cascades', {\it Journal of
Political Economy}, 100(5), 992--1026.

Blanchard O. \& M.W. Watson, (1982), `Bubbles, Rational Expectations and
Financial Markets', P.~Wachtel (Ed.), in {\it Crises in the Economic and
Financial Structure}, Lexington Books, 295--315.

Bollerslev T., R.Y. Chou, N. Jayaraman \& K.F. Kroner (1991), `Les 
mod\`eles ARCH
en finance : un point sur la th\'eorie et les r\'esultats 
empiriques', {\it Annales
d'Economie et de Statistiques}, 24, Oct.

Bollerslev T. (1987), `A Conditional Heteroskedastic Time Series Model for
Speculative Prices and Rates of Return', {\it Review of Economics and
Statistics}, 69.

Bouchaud, J.-P. and R. Cont (1998),
A Langevin approach to stock market fluctuations and crashes,
{\it Eur. Phys. J. B}, 6, 543--550.

Brock W.A. (1988), `Nonlinearity and Complex Dynamics in Economics and
Finance', in P.W. Anderson, K.J. Arrow, D. Pines (Eds.), {\it The Economy as an
Evolving Complex System}, Addison-Wesley.

Brock W.A. \& W.D. Dechert (1988), `Theorems on Distinguishing Deterministic
from Random Systems', in Barnett W.A., E.R. Berndt \& H. White (Eds.), {\it
Dynamic Econometric Modeling}, Chap.~12, 247--265, Cambridge, 
Cambridge University
Press.

Brock W.A., W.D. Dechert \& J. Scheinkman (1987), `A Test for Independence
Based on the Correlation Dimension', Working Paper, University of Wisconsin at
Madison, University of Houston and University of Chicago.

Brock W.A., D. Hsieh \& B. LeBaron (1991), {\it Nonlinear Dynamics, Chaos and
Instability : Statistical Theory and Economic Evidence}, Cambridge, MAD Press.

Campbell, J.Y., A.W. Lo, A.C. MacKinlay (1997),
{\it The econometrics of financial markets} (Princeton, N.J. : Princeton
University Press).

Challet, D. and Zhang, Y.C. (1997),
Emergence of cooperation and organization in an evolutionary game,
{\it Physica A}, 246, 407--18.

Challet, D., Marsili, M. and Zecchina, R. (2000),
Statistical mechanics of systems with heterogeneous agents: Minority
games, {\it Physical Review Letters}, 84, 1824--1827.

Challet, D., Marsili, M. and Zhang, Y.C. (2000),
Modeling market mechanism with minority game, {\it Physica A},
276, 284--315.

Collet P. \& J.-P. Eckmann (1980), {\it Iterated Maps on the Interval as
Dynamical Systems}, Boston, Birkh\"auser.

Cont., R. and Bouchaud, J.-P. (2000)
Herd behavior and aggregate fluctuations in financial markets, Macroeconomic
Dynamics 4, 170-196.

Corcos A. (1993), {\it Bruit et Chaos sur les march\'es financiers}, Th\`ese de
Doctorat, Universit\'e Panth\'eon-Assas, Paris.

Dacorogna, M.M. ,  U.A. M\"uller, R.B. Olsen, O.V. Pictet (1998),
Modelling Short-Term Volatility with GARCH and HARCH Models,
in ``Nonlinear Modelling of High Frequency Financial Time Series,'' by C.
Dunis, B. Zhou (John Wiley \& Sons).

Day R.H. (1982), `Irregular Growth Cycles', {\it American Economic Review},
June, 72, 406--414.

Day R.H. (1983), `The Emergence of Chaos from Classical Economic Growth', {\it
Quarterly Journal of Economics}, May, 48, 201--213.

De Bondt, W.F.M. and Thaler, R.H. (1995),
Financial decision-making in markets and firms: a behavioral
perspective, in {\it Finance}, R.A. Jarrow, V. Maksimovic, W.T. Ziemba, eds.,
Handbooks in Operations Research
and Management Science {\bf 9}, 385-410 (Elsevier Science, Amsterdam;
New York).

Ding, Z., Granger, C.W.J. and Engle, R. (1993),
A long memory property of stock returns and a new model, {\it Journal 
of Empirical Finance}, 1,
83--106.

Ding, Z., Granger, C.W.J. (1996),
Modeling volatility persistence of speculative returns: A new approach,
{\it Journal of Econometrics}, 73, 185--215.

Eckmann J.-P. (1981), `Roads to Turbulence in Dissipative Dynamical 
Systems', {\it
Reviews of Modern Physics}  53, 643--654. Reprinted in Universality in Chaos,
P.~Cvitanovi\'c (Ed.), Adam Hilger, 94 (1984).

Eckmann J.-P., S. Oliffson Kamphorst, D. Ruelle \& J. Scheinkman (1988),
`Lyapunov Exponents for Stock Returns', in P.W. Anderson, K.J. Arrow, D. Pines
(Eds.), {\it The Economy as an Evolving Complex System}, Addison-Wesley.

Eckmann J.-P. \& D. Ruelle (1985), `Ergodic Theory of Chaos and Strange
Attractors', {\it Reviews of Modern Physics}, 57, 617--656.

Eckmann J.-P. \& D. Ruelle (1992), `Fundamental Limitations for Estimating
Dimensions and Lyapunov Exponents in Dynamical Systems', {\it Physica} D56,
185--187.

Egenter, E., T. Lux and D. Stauffer (1999),
Finite-size effects in Monte Carlo simulations of two stock market models,
{\it Physica A}, 268, 250--256.

Engle R.F. (1982), `Autoregressive Conditional Heteroskedasticity with
Estimates of the Variance of U.K. Inflation', {\it Econometrica}, 50.

Fama E. (1965), `The Behavior of Stock Market Prices', {\it Journal of
Business}, 38, 34--105.

Farmer, J.D. (1998), Market force, ecology and evolution, preprint
available at adap-org/9812005

Farmer, J.D. and S. Joshi (2001), The price
dynamics of common strategies, to appear in the Journal of Economic
Behavior and Organization, e-print at

Feller W. (1966), {\it An Introduction to Probability Theory and its
Applications}, Vol.~II, New York, John Wiley.

Goldberg, J. and von Nitzsch, R. (translated
Morris, A.) (2001) {\it Behavioral finance} (Chichester, England; New 
York: John Wiley).

Graham B. and D.L. Dodd (1934), {\it Security analysis}, 1st edition 
(McGraw-Hill).

Grandmont J.-P. \& P. Malgrange (1986), `Nonlinear Economic Dynamics :
Introduction', {\it Journal of Economic Theory}, 40, Oct, 3--11.

Grandmont J.-P. (1985), `On Endogenous Competitive Business Cycles', {\it
Econometrica}, 53.

Grandmont J.-P. (1987), {\it Nonlinear Economic Dynamics}, New
York, Academic Press.

Grauwe De P., H. Dewachter \& M. Embrechts (1993), {\it Foreign Exchange
Models}, Oxford, Blackwell.

Grauwe De P., K. Vansanten (1990), `Deterministic Chaos in the Foreign Exchange
Market', Working Paper, CEPR, Katholieke Universiteit Leuven, Belgium.

Grossman S.J. (1977), `The Existence of Future Markets, Noisy Rational
Expectations and Informational Externalities', {\it Review of Economic
Studies}, 64.

Grossman S.J. (1981), `An Introduction to the Theory of Rational Expectations
under Asymmetric Information', {\it Review of Economic Studies}, 154.

Grossman S.J. \& R.J. Shiller (1981), `The Determinants of the Variability of
Stock Market Prices', {\it American Economic Review}, 71(2), 222--227.

Grossman S.J. \& J. Stiglitz (1980), `The Impossibility of Informationally
Efficient Markets', {\it American Economic Review}, 70, June.

Hsieh D. \& B. LeBaron (1988) `Finite Sample Properties of the BDS Statistics,
Working Paper', University of Chicago and University of Wisconsin.

Hsieh D. (1989), `Testing for Nonlinearity in Daily Foreign Exchange Rate
Changes', {\it Journal of Business}, 62.

Hsieh D. (1991), `Chaos and Nonlinear Dynamics : Application to Financial
Markets', {\it The Journal of Finance}, Dec.

Hsieh D. (1992), `Implications of Nonlinear Dynamics for Financial Risk
Management', Workshop on Nonlinear Dynamics in Economics, European University
Institute, July.
 
Huang, Z.-F. and S. Solomon (2001),
Finite market size as a source of extreme wealth
inequality and market instability, {\it Physica A}, 294, 503--513.

Ide, K. and D. Sornette (2001),
Oscillatory Finite-Time Singularities in Finance, Population and Rupture,
preprint  (http://arXiv.org/abs/cond-mat/0106047)

Jefferies, P., Hart, M.L., Hui, P.M. and Johnson, N.F. (2001),
 From market games to real-world markets, {\it European Physical Journal B},
20, 493--501.

Jevons S. (1871), {\it Theory of Political Economy}, Pelican Classics.

Johansen, A. and D. Sornette (1999),
Critical Crashes, {\it RISK}, 12 (1), 91--94.

Johansen, A., D. Sornette and O. Ledoit (1999),
Predicting Financial Crashes using discrete scale invariance,
{\it Journal of Risk}, 1 (4), 5--32.

Johansen, A., O. Ledoit and D. Sornette (2000),
Crashes as critical points,  {\it International Journal of 
Theoretical and Applied Finance},
3 (2),  219--255.

Johansen, A. and D. Sornette (2000),
The Nasdaq crash of April 2000: Yet another example of
log-periodicity in a speculative bubble ending in a crash,
{\it European Physical Journal B}, 17, 319--328.

Johansen, A. and D. Sornette (2001)
Large Stock Market Price Drawdowns Are Outliers, in press in {\it 
Journal of Risk}
(http://arXiv.org/abs/cond-mat/0010050)

Keynes J.M. (1936), {\it The General Theory of Employment, Interest and Money},
London, McMillan.

Kim, G.W. and H.M. Markowitz  (1989), {\it J. Portfolio Management}, 16, 45.

Kreps D. (1977), `Fulfilled Expectations Equilibria', {\it Journal of Economic
Theory}, 14.

Laherr\`ere, J. and D. Sornette (1998),
Stretched exponential distributions in Nature and Economy: ``Fat tails''
with characteristic scales, {\it European Physical Journal}, B 2, 525--539.

LeBaron B. (1988), `The Changing Structure of Stock Returns', Working Paper,
University of Wisconsin.

Levy, M. H. Levy and S. Solomon (1995),
Microscopic simulation of the stock market -- the effect of microscopic
diversity, {\it J. Physique I}, 5,  1087--1107,

Levy, M., H. Levy and S. Solomon (2000),
{\it The microscopic simulation of financial markets: from investor behavior
to market phenomena} (Academic Press, San Diego).

Lo, A.W. and A.C.  MacKinlay (1999),
{\it A Non-Random Walk down Wall Street} (Princeton University Press).

Lorenz E. (1963), `Deterministic Nonperiodic Flow', {\it Journal of Atmospheric
Sciences}, 20.

Lux, L. (1996), The stable Paretian hypothesis and the frequency of large
returns: an examination of major German stocks, {\it
Appl. Financial Economics}, 6, 463--475.

Lux, T. and M. Marchesi (1999), Scaling and criticality
in a stochastic multi-agent model of a financial market, {\it Nature} 
297, 498--500.

Manneville P. (1991), {\it Structures dissipatives, chaos et turbulence}, CEA,
Saclay.

Maug, E. and Naik, N. (1995), Herding and delegated portfolio management: The
impact of relative performance evaluation on asset allocation, 
Working paper, Duke University.

May R. (1976), `Simple Mathematical Models with Very Complicated 
Dynamics', {\it
Nature}, 261.

M\"uller, U.A., M.M. Dacorogna, R. Dav\'e, R.B. Olsen, O.V. Pictet and J.E. von
Weizs\"acker (1997),
Volatilities of Different Time Resolutions - Analyzing the Dynamics of Market
Components, {\it Journal of Empirical Finance}, 4, 213--240.

Muth J. (1961), `Rational Expectations and the Theory of Price Movements', {\it
Econometrica}, July.

Muzy, J.-F., D. Sornette, J. Delour and A.~Arneodo (2001),
Multifractal returns and Hierarchical Portfolio Theory,
{\it Quantitative Finance}, 1 (1), 131--s148.

Orl\'ean A. (1986), `Mim\'etisme et anticipations rationnelles : perspectives
keyn\'esiennes', {\it Recherches Economiques de Louvain}, 52(1), 45--66.

Orl\'ean A. (1989), `Comportements mim\'etiques et diversit\'e des 
opinions sur les
march\'es financiers', in Bourguinat H. \& Artus P. (Eds.), {\it 
Th\'eorie Economique
et Crise des March\'es financiers}, Economica, 45--65.

Orl\'ean A. (1990), `Le r\^ole des influences interpersonnelles dans la
d\'etermination des cours boursiers', {\it Revue Economique}, 41(5), 839--868.

Orl\'ean A. (1992), `Contagion des opinions et fonctionnement des march\'es
financiers', {\it Revue Economique}, 43(3), 685--697.

Pagan, A. (1996), The Econometrics of Financial Markets,
{\it Journal of Empirical Finance}, 3, 15--102.

Plerou, V., Gopikrishnan, P., Amaral, L.A.N., Meyer, M. \& Stanley, 
H.E. (1999),
Scaling of distribution of price fluctuations of individual companies,
{\it Physical Review E}, 60, 6519--6529.

Radner R. (1972), `Existence of Equilibrium of Plans, Prices and Price
Expectations in a Sequence of Markets', {\it Econometrica}, 40.

Radner R. (1979), `Rational Expectations Equilibrium : Generic Existence and
the Information Revealed by Price', {\it Econometrica}, 47.

Roehner, B.M. and D. Sornette (1998),
The sharp peak-flat trough pattern and critical speculation,
{\it European Physical Journal B}, 4, 387--399.

Scharfstein, D. and Stein, J. (1990)
Herd behavior and investment, American Economic Review {\bf 80}, 465--479.

Scheinkman J.A. \& B. LeBaron (1989a), `Nonlinear Dynamics and GNP Data', in
Barnett W.A., J.~Geweke \& K. Shell (Eds.), {\it Economic Complexity : Chaos,
Sunspots, Bubbles and Nonlinearity}, Cambridge University Press.

Scheinkman J.A. \& B. LeBaron (1989b), `Nonlinear Dynamics and Stock Returns',
{\it Journal of Business}, 62(3).

Shefrin, H. (2000),
{\it Beyond greed and fear: understanding behavioral finance and the
psychology of investing} (Boston, Mass.: Harvard Business School Press).

Shiller R. (1981), `Do Stock Prices Move too Much to Be Justified by Subsequent
Changes in Dividends?', {\it American Economic Review}, 71(3), June.

Shiller R. (1984), `Stock Prices and Social Dynamics', {\it Brookings Papers on
Economic Activity}, 457--498.

Shiller, R.J. (2000),
{\it Irrational exuberance} (Princeton University Press, Princeton, NJ).

Shleifer, A. (2000),
{\it Inefficient markets: an introduction to behavioral finance}
(New York: Oxford University Press).

Sornette, D. (2001), {\it Critical market crashes} (Princeton University Press,
Princeton, NJ) in press.

Sornette, D. and A. Johansen (2001),
Significance of log-periodic precursors to financial crashes,
{\it Quantitative Finance}, 1 (4), 452--471.

Sornette, D. and J.V. Andersen (2001),
A Nonlinear Super-Exponential Rational Model of Speculative Financial Bubbles,
preprint at http://arXiv.org/abs/cond-mat/0104341

Sornette, D. and K. Ide  (2001),
Theory of self-similar oscillatory finite-time singularities
in Finance, Population and Rupture, preprint
(http://arXiv.org/abs/cond-mat/0106054)

Sornette, D., A. Johansen and J.-P. Bouchaud (1996),
Stock market crashes, Precursors and Replicas,
{\it J.Phys.I France}, 6, 167--175.

Stauffer, D. (1999),
Finite-Size Effects in Lux-Marchesi and Other Microscopic Market Models,
Genoa economics meeting, June 1999 (electronic distribution only, edited by
M. Marchesi; dibe.unige.it/wehia).

Stutzer M. (1980), `Chaotic Dynamics and Bifurcation in a Macro-Model', {\it
Journal of Economic Dynamics and Control}.

Thaler, R.H., ed. (1993),
{\it Advances in behavioral finance} (New York: Russell Sage Foundation)

Topol R. (1991), `Bubbles and Volatility of Stock Prices: Effect of Mimetic
Contagion', {\it Economic Journal}, 101, 786--800.

Trueman, B. (1994), Analyst forecasts and herding behavior,
{\it The Review of Financial Studies}, 7, 97--124.

Van der Ploeg F. (1986), `Rational Expectations, Risk and Chaos in Financial
Markets', {\it Economic Journal}, 96, Supplement.

Welch, I. (1992),
Sequential sales, learning, and cascades, {\it Journal of Finance}, 
47, 695--732.
see also http://welch.som.yale.edu/cascades for
an annotated bibliography and resource reference on ``information cascades''.

West K.D. (1988), `Bubbles, Fads and Stock Price Volatility Tests: a 
Partial Evaluation',
{\it Journal of Finance}, 43(3), 639--655.

Zwiebel, J. (1995), Corporate conservatism and
relative compensation, {\it Journal of Political Economy}, 103, 1--25.

\newpage

\newpage

\newpage

\newpage

\newpage

\newpage

\newpage

\newpage

\newpage

\newpage

\newpage

\newpage

\newpage

\newpage

\newpage

\clearpage

\clearpage

\clearpage

\end{document}